\documentclass[prd,preprint
,superscriptaddress,showkeys,nofootinbib
,tightenlines,colorlinks=true,citecolor=blue
]{revtex4}
\usepackage[utf8]{inputenc}
\usepackage{mathrsfs}
\usepackage{latexsym,bm}
\usepackage{graphicx}
\usepackage{indentfirst}
\usepackage{slashed}
\usepackage{amsmath}
\usepackage{amssymb}
\usepackage{color}
\usepackage{hyperref}
\usepackage{epsfig}
\usepackage[titletoc]{appendix}
\usepackage{multirow}%
\usepackage{rotating}
\usepackage{epstopdf}
\usepackage{extarrows}
\usepackage{tabularx}
\usepackage{soul} 
\usepackage{xcolor}
\usepackage{lipsum}
\usepackage{hyperref}

\newcommand{\bea}{\begin{eqnarray}} 
\newcommand{\eea}{\end{eqnarray}}

\newcommand{\al}{&\!\!\!\!}
\newcommand{\lag}{\mathcal{L}}

\newcommand{\itp}{\affiliation{CAS Key Laboratory of Theoretical Physics,
            Institute of Theoretical Physics,\\ Chinese Academy of Sciences,
            Beijing 100190, China}}
\newcommand{\bonn}{\affiliation{Helmholtz-Institut f\"ur Strahlen- und
             Kernphysik and Bethe Center for Theoretical Physics,\\
             Universit\"at Bonn,  D-53115 Bonn, Germany}}
\newcommand{\fzj}{\affiliation{Institute for
           Advanced Simulation, Institut f\"ur Kernphysik and
           J\"ulich Center for Hadron Physics,\\
           Forschungszentrum J\"ulich, D-52425 J\"ulich, Germany}}

\newcommand{\ucas}{\affiliation{School of Physical Sciences,
            University of Chinese Academy of Sciences,\\
            Beijing 100049, China}}

\newcommand{\tbilisi}{\affiliation{Tbilisi State University, 0186 Tbilisi, Georgia}}

\begin{document}

\title{Implications of chiral symmetry on $\bm{S}$-wave pionic resonances and the scalar charmed mesons}

\author{Meng-Lin Du}
\email{du@hiskp.uni-bonn.de}
\bonn

\author{Feng-Kun~Guo}
\email{fkguo@itp.ac.cn}
\itp\ucas

\author{Ulf-G.~Mei{\ss}ner} 
\email{meissner@hiskp.uni-bonn.de}
\bonn\fzj\tbilisi

\begin{abstract}

The chiral symmetry of QCD requires energy-dependent pionic strong interactions at low energies. This constraint,
however, is not fulfilled by the usual Breit--Wigner parameterization of pionic resonances, leading to
masses larger than the real ones. We derive relations between nonleptonic three-body decays of the $B$-meson
into a $D$-meson and a pair of light pseudoscalar mesons based on SU(3) chiral symmetry. Employing effective
field theory methods, we demonstrate that taking into account the final-state interactions, the experimental
data of the decays $B^-\to D^+\pi^-\pi^-$, $B_s^0\to \bar{D}^0K^-\pi^+$, $B^0\to\bar{D}^0\pi^-\pi^+$,
$B^-\to D^+\pi^-K^-$ and $B^0\to\bar{D}^0\pi^-K^+$ can all be described by the nonperturbative $\pi/\eta/K$-$D/D_s$
scattering amplitudes previously obtained from a combination of chiral effective field theory and
lattice QCD calculations.  The results provide a strong support of the scenario that the broad scalar charmed
meson $D^\ast_0(2400)$ should be replaced by two states, the lower one of which has a mass of around 2.1~GeV,
much smaller than that extracted from experimental data using a Breit--Wigner parameterization.

\end{abstract}

\pacs{12.39.Fe, 13.75.Lb, 14.40.Lb}
\keywords{Chiral symmetry, unitarity, coupled-channel dynamics, nonleptonic $B$-meson decays}

\maketitle
\newpage

\section{Introduction}

Solving Quantum Chromodynamics (QCD), the fundamental theory of the strong interaction, is one of the most urgent and
challenging tasks in contemporary physics. In principle, the hadronic spectrum is determined by QCD. However, the
high nonlinearity of QCD at low energies does not allow for analytical solutions. A direct and clear connection
between the fundamental theory and the observed hadronic spectrum is so far absent. Studies of the hadronic spectrum
are particularly important in testing the fundamental theory. Therefore, establishing the correct spectroscopy plays
an essential role for understanding how the fundamental theory, whose  degrees of freedom are quarks and gluons,
produces the hadronic spectrum.

The hadronic spectrum has received renewed interest with the recently collected vast amounts of experimental
data for hadronic processes. In particular, many resonant structures were observed, and they could be the so-called exotic states which cannot be accommodated in the conventional quark model.
The lightest scalar ($J^P=0^+$) charm-strange meson $D_{s0}^\ast(2317)$~\cite{Aubert:2003fg} and the axial-vector
($J^P=1^+$) one $D_{s1}(2460)$~\cite{Besson:2003cp} have attracted much attention as they are significantly
lighter than the predictions from the quark model calculations, which are around 2.48~GeV and 2.55~GeV~\cite{Godfrey:1985xj,Godfrey:2015dva,Ebert:2009ua}, respectively. It is also noticed that the mass difference
between the two states is equal to that between the ground-state pseudoscalar $D^+$ and the vector $D^{\ast +}$
within 2~MeV. Since attempts to adjust the quark model to adapt the two new states were at odds with previous
expectations and raised new puzzles~\cite{Cahn:2003cw}, various theoretical investigations were
stimulated to unravel their nature. Model proposed for them include $D^{(\ast)}K$ molecules~\cite{Barnes:2003dj,Szczepaniak:2003vy,Chen:2004dy,Guo:2006fu,Guo:2006rp}, tetraquarks~\cite{Cheng:2003kg,Maiani:2004vq},
a mixture of $c\bar{q}$ with tetraquarks~\cite{Browder:2003fk} and a cousin of the light scalar mesons~\cite{vanBeveren:2003kd}.
The situation has become more obscure by the subsequent observations of the scalar and axial-vector charm nonstrange mesons,
$D_0^\ast(2400)$~\cite{Abe:2003zm,Link:2003bd} and $D_1(2430)$~\cite{Abe:2003zm} in 2004. Their quantum
numbers indicate that they should be the SU(3) partners of the $D_{s0}^\ast (2317)$ and $D_{s1}(2460)$, respectively,
which sets the starting point for various theoretical studies, see, e.g., Refs.~\cite{Mehen:2005hc,Ananthanarayan:2007sa,Colangelo:2012xi}. However, this assignment immediately raises a puzzle: Why are the masses of the two strange mesons,
$D_{s0}^\ast(2317)$ and $D_{s1}(2460)$, almost equal to or even lower than their nonstrange siblings, i.e.,
the $D_0^\ast(2400)$ and $D_1(2430)$.

While the mass of $D_{s0}^\ast (2317)$ is well measured and its width is very narrow~\cite{Tanabashi:2018oca},
the situation for the $D_0^\ast(2400)$ is obscure. Its width is very broad and its mass has a significant spread
in different experimental measurements, e.g., from $2297\pm 8\pm 5\pm 19$~MeV at BABAR~\cite{Aubert:2009wg} to
$2403 \pm 14 \pm 35$~MeV at FOCUS~\cite{Link:2003bd}. In the potential quark model, it would correspond to the lowest $c\bar{q}$ $P$-wave state with
a predicted mass of about $2.4$~GeV~\cite{Godfrey:1985xj,Godfrey:2015dva}, which is larger than the averaged mass $2318\pm 29$~MeV by the Particle Data Group (PDG)~\cite{Tanabashi:2018oca}. One reason why the analyses that led to the resonances parameters of the $D_0^\ast(2400)$
and $D_1(2430)$ in Refs.~\cite{Link:2003bd,Abe:2003zm,Aubert:2009wg,Aaij:2015kqa,Aaij:2015sqa}, used in the PDG average, should be questioned is
that the amplitudes used were inconsistent with the constraints from the chiral symmetry of QCD, which, as a consequence
of the Goldstone theorem, requires energy-dependent couplings for the pionic couplings. The standard Breit--Wigner (BW) parameterizations
used in the experimental analyses correspond to using constant vertices, and lead to  masses larger than their real values.
Moreover, the signal ranges for these two states overlap with higher $S$-wave thresholds ($D^{(*)}\eta,D_s^{(*)}\bar K$) that need to be considered
in a sound analysis. Fortunately, a theoretical framework satisfying such requirements is provided by
the chiral perturbation theory (ChPT) for charmed mesons extended with a unitarization procedure, see, e.g.,
Refs.~\cite{Kolomeitsev:2003ac,Hofmann:2003je,Gamermann:2006nm,Guo:2006fu,Guo:2006rp,Guo:2008gp,Guo:2009ct,Liu:2012zya,Yao:2015qia,Du:2017ttu,Guo:2018kno,Guo:2018tjx}. 

In Ref.~\cite{Du:2017zvv} it is demonstrated that in the scenario that the lightest scalar (axial-vector) charmed mesons owe their existence to the nonperturbative $\pi/\eta/K$-$D^{(\ast)}/D_s^{(\ast)}$ scattering, all of the above mentioned puzzles
in  charmed meson spectroscopy get resolved. The $D_{s0}^\ast(2317)$ and $D_{s1}(2460)$ are dominantly $DK$
and $D^\ast K$ hadronic molecules (reviewed in Ref.~\cite{Guo:2017jvc}), respectively; heavy-quark spin symmetry predicts that their binding energies
are independent of the heavy meson spin. Most importantly, the ordering of the lightest strange, i.e.
$D_{s0}^\ast(2317)$, and the non-strange scalar, i.e. $D_0^\ast(2400)$, becomes natural. The broad $D_0^\ast(2400)$
listed in the Review of Particle Physics (RPP)~\cite{Tanabashi:2018oca} should be replaced by two states with
the lighter one located more than 100~MeV below its strange sibling, i.e. $D_{s0}^\ast(2317)$~\cite{Albaladejo:2016lbb,Du:2017zvv}, as well as the masses of $D_0^\ast(2400)$ listed in RPP~\cite{Tanabashi:2018oca}, which were extracted from BW parameterizations. Furthermore, it was demonstrated
that the amplitudes for Goldstone boson ($\phi$)-$D^{(\ast)}$ scattering determined by combining unitarized ChPT
with lattice QCD in Ref.~\cite{Liu:2012zya} are fully consistent  the recent high quality LHCb data of the angular moments for the 
decay $B^-\to D^+\pi^-\pi^-$~\cite{Aaij:2016fma}. In particular, that the LHCb data can be fitted with an amplitude having two $D_0^*$ states, none of whose masses agree with that of the experimental extractions for the $D_0^*(2400)$, means that the resonance parameters for the scalar charm-nonstrange meson need to be reconsidered by the PDG and by the community.

As the final-state interaction (FSI) between the two $\pi^-$ is negligible since they are in an isospin-two state, the reaction $B^-\to D^+\pi^-\pi^-$~\cite{Aaij:2016fma} provides a great access to
the $D\pi$ system and thus to the charm-nonstrange mesons. The energy range of the $D^\ast_0(2400)$ overlaps
with two $S$-wave thresholds ($D\eta$ and $D_s\bar K$), and thus these channels need to be considered and leave their imprints on observables~\cite{Du:2017zvv}. It means that all channels, i.e., $D^+\pi^-$, $D^0\pi^0$, $D^0\eta$ and $D_s^+K^-$,
coupled to the $D^+\pi^-$ need to be taken into account in the intermediate states for describing the $D^+\pi^-$ distributions $\lesssim 2.5$~GeV of the decay $B^-\to D^+\pi^-\pi^-$. The $D^0$, $D^+$, and $D_s^+$
mesons form a $\bar{3}$ representation of the light-quark SU(3) flavor symmetry. Using the SU(3) transformation properties of the effective
Hamiltonian for weak nonleptonic $B$-meson three-body decays, a ratio between the four channels can be derived~\cite{Savage:1989ub,Du:2017zvv}.
Because the two $\pi^-$ could only be in a state where the relative angular momentum is even, the effective
Lagrangian for the weak decays of $\bar{B}$ to $D$, induced by the Cabibbo-allowed transition $b\to c\bar{u}d$,
with the emission of two symmetrized light pseudoscalar mesons was constructed in the Appendix of
Ref.~\cite{Du:2017zvv}. In this paper, we extend the effective weak Lagrangian to include the
Cabibbo-suppressed decays with two nonidentical light pseudoscalar mesons. Since we will focus on the energy region only a few hundred MeV above the corresponding thresholds, one of the light pseudoscalar mesons would move fast. For simplicity, we do not consider
the FSIs from crossed channels, i.e., the interaction between $D/D_s$ and the fast moving
light pseudoscalar meson and the interaction between the two light pseudoscalar mesons. As in the isobar model, the
crossed-channel effects are simply assumed to be encoded in an extra complex factor which would be fixed by
fitting to experimental data. However, the final-state interaction between the charmed meson and the soft
light pseudoscalar meson and that from coupled channels are taken into account. 

While the weak production vertices can be derived from the weak effective Lagrangian, the rescatterings
between the charmed mesons and the soft light pseudoscalar mesons are described by the unitarized ChPT,
see e.g. in Refs.~\cite{Guo:2009ct,Liu:2012zya,Yao:2015qia,Du:2017ttu,Guo:2018tjx}. In Ref.~\cite{Liu:2012zya}, the
low-energy constants (LECs) of ChPT for charmed mesons, and thus the chiral amplitudes for the $D$-$\phi$ scattering,
are obtained by fitting to a lattice QCD calculation of the scattering lengths for the channels of
with connected Wick contractions only, i.e. $D\pi$ with isospin $I=3/2$, $D\bar{K}$ with $I=0$ and 1, $D_sK$ and $D_s\pi$. With the parameters fixed in Ref.~\cite{Liu:2012zya}, the predicted energy levels for the
coupled-channel systems in a finite volume are found~\cite{Albaladejo:2016lbb,Albaladejo:2018mhb} to be in a remarkable agreement with recent lattice QCD
calculations~\cite{Moir:2016srx,Bali:2017pdv}.\footnote{In an updated fit~\cite{Guo:2018tjx} to the lattice data, several different fits were found, and the parameters in the set which can reproduce the $D_{s0}^*(2317)$ mass as well as satisfy the $N_c$ scaling are similar to those determined in Ref.~\cite{Liu:2012zya}.} It indicates the reliability of 
the well constrained chiral amplitudes in Ref.~\cite{Liu:2012zya}.
As in Ref.~\cite{Du:2017zvv}, we will demonstrate that the amplitudes obtained in Ref.~\cite{Liu:2012zya}
are consistent with the experimental data on the three-body decays $B_s^0\to \bar{D}^0K^-\pi^+$,
$B^0\to\bar{D}^0\pi^-\pi^+$, $B^-\to D^+\pi^-K^-$ and $B^0\to\bar{D}^0\pi^-K^+$, and try to determine the LECs
of the weak effective Lagrangian from these decays.

This paper is organized as follows. In section~\ref{sec:BW}, we will briefly discuss the implication of the
chiral symmetry on pionic BW resonances. In section~\ref{sec:eft}, the chiral effective weak Lagrangian
describing the $B$ meson decays to the $D$ meson associated with two light pseudoscalar mesons is constructed. The decay
amplitudes satisfying  unitarity for the reactions $B_s^0\to \bar{D}^0K^-\pi^+$, $B^0\to\bar{D}^0\pi^-\pi^+$,
$B^-\to D^+\pi^-K^-$ and $B^0\to\bar{D}^0\pi^-K^+$ are derived in section~\ref{sec:amp}. In section~\ref{sec:fit},
the so-called angular moments are calculated and fitted to experimental data. Finally, section~\ref{sec:sum}
comprises a brief summary.

\section{Implication of chiral symmetry on pionic resonances}\label{sec:BW}

The small masses of the $u$-, $d$- and $s$-quark compared with the QCD scale $\Lambda_\text{QCD}$ induce
an approximate SU(3)$_L\times$SU(3)$_R$ chiral symmetry for the strong interactions. As the chiral symmetry
is spontaneously broken to its diagonal subgroup SU(3)$_V$, the $\pi$, $K$, and $\eta$ mesons arise as
the pseudo-Goldstone bosons. At the leading order (LO) in the chiral expansion, the interactions of Goldstone bosons with themselves and with matter fields are of a derivative form and thus energy-dependent. However, the standard BW parameterization used in the experimental analyses to extract the resonance parameters (mass and width)
corresponds to a constant interaction for $S$-wave vertices. Were the energy-dependent interaction induced by the derivative coupling considered,
the mass fitted using a normal BW parameterization would be shifted to a lower value. This point is illustrated in the following.

For simplicity, we neglect the energy-dependence
of the decay width in this section. The BW parameterization for a resonance in the $D\pi$ $S$-wave reads
\bea\label{eq:bwori}
F_0(s) \propto \dfrac{1}{s-m_0^2+im_0\Gamma},
\eea
where the $m_0$ and $\Gamma$ represent the BW mass and width of the resonance, respectively. Now let us calculate the
peak position of this BW parameterization. It corresponds to the energy where $$\dfrac{d}{ds}|F_0(s)|^2
\propto -\dfrac{2(s-m_0^2)}{\big[(s-m_0^2)^2+m_0^2\Gamma^2\big]^2}=0,$$ 
which means that the BW mass for a
resonance corresponds to the value of the peak position. In order to take the chiral symmetric constraint into account, one may simply
modify Eq.~\eqref{eq:bwori} by introducing a pion energy factor as
\bea\label{eq:bwrev}
F_0^\prime(s)\propto \dfrac{E_\pi}{s-m_0^2+i m_0\Gamma}
\eea
with $E_\pi=(s+M_\pi^2-M_D^2)/(2\sqrt{s})$ the energy of the produced soft pion in the rest frame of the $D\pi$ system.
The peak position $s_\text{peak}$ can be obtained with the same approach, that is
\bea
\dfrac{d}{ds}|F_0^\prime(s)|^2 \Big\arrowvert_{s=s_\text{peak}} = 0.
\eea
It is easy to see that the peak position $s_\text{peak}$ is shifted from $m_0^2$. The shift is expected to be
small compared with $m_0$ for $\Gamma\ll m_0$. Let $s_\text{peak}=(m_0+\Delta)^2$. Keeping only  the
term linear in $\Delta$, one gets
\bea
\Delta \simeq \dfrac{\Gamma^2(m_0^2 -M_\pi^2+M_D^2)}{2m_0\big[ 2(m_0^2+M_\pi^2-M_D^2) 
-\Gamma^2\big]} =\dfrac{\Gamma^2 E_D}{4m_0E_\pi-\Gamma^2},
\eea
where $E_D$ is the energy of the produced $D$ in the rest frame of $D\pi$ system with a total energy $m_0$. Thus, for the
case $4m_0E_\pi >  \Gamma^2$, e.g., that of the $D_0^\ast(2400)$, the shift $\Delta$ is positive, which means that the
mass of the resonance is lower than the peak position. 

Nevertheless, the modification in Eq.~\eqref{eq:bwrev} can only be applied in a small energy region
before the coupled-channel effect becomes important, and thus is not  practical or systematic. A theoretical framework
satisfying both the chiral symmetry constraint and unitarity taking into account coupled channels is needed. Such requirements are fulfilled by the unitarized ChPT, see, e.g.,
Refs.~\cite{Dobado:1996ps,Oller:1997ti,Oller:1998hw,Oller:2000fj,GomezNicola:2001as}.

\section{Chiral effective Lagrangian}\label{sec:eft}

While the ground state pseudoscalar octet ($\phi$) at low energies can be treated as the pseudo-Goldstone bosons associated
with the spontaneous breaking of the approximate chiral symmetry into its diagonal subgroup SU(3)$_V$, SU(3)$_V$ is realized in the Wigner--Weyl mode. It means that the non-Goldstone bosons
(matter fields) form multiplets which correspond to different irreducible representations of SU(3).
In particular, the $B^-$, $\bar{B}^0$, and $\bar{B}_s^0$ mesons, and $D^0$, $D^+$ and $D^+_s$ mesons
form two $\bar{3}$ fundamental representations of SU(3). In the language of the Callan--Coleman--Wess--Zumino formalism~\cite{Coleman:1969sm,Callan:1969sn}, the Goldstone bosons are realized nonlinearly in $u(x)$,
which transforms under a global chiral transformation $g_L\times g_R\in \text{SU(3)}_L\times \text{SU(3)}_R$ as 
$$u(x)\mapsto g_R u(x) h(x)^\dag = h(x)u(x)g_L^\dag,$$
where the compensator field $h(x)$ is a nonlinear function of $g_L$, $g_R$ and $u(x)$, and it reduces to $h(x)=g_L=g_R$, independent of
$u(x)$, for a SU(3)$_V$ transformation. We parameterize $u$ as $u=\exp (i\phi/(\sqrt{2}F_0))$, where
$F_0$ is the pion decay constant in the chiral limit and 
\bea
\phi =\begin{pmatrix}
\frac{1}{\sqrt{2}}\pi^0+\frac{1}{\sqrt{6}}\eta  & \pi^+ & K^+  \\
\pi^- & -\frac{1}{\sqrt{2}}\pi^0 +\frac{1}{\sqrt{6}}\eta & K^0 \\
K^- & \bar{K}^0 & -\frac{2}{\sqrt{6}}\eta
\end{pmatrix}.
\eea
From the chiral transformation properties of $u(x)$, one can derive the axial-vector $u_\mu$ and
vector $\Gamma_\mu$ currents~\cite{Coleman:1969sm,Callan:1969sn}:\footnote{The inclusion of the  external currents is straightforward, see  Refs.~\cite{Gasser:1983yg,Gasser:1984gg}.}
\bea
u_\mu \al = \al i(u^\dag \partial_\mu u -u \partial_\mu u^\dag), \nonumber\\
\Gamma_\mu \al =\al \frac{1}{2}(u^\dag \partial_\mu u +u \partial_\mu u^\dag ).
\eea
The axial-vector $u_\mu$ transforms homogeneously,
\bea
u_\mu \mapsto h u_\mu h^\dag,
\eea
whereas $\Gamma_\mu$ transforms inhomogeneously,
\bea
\Gamma_\mu \mapsto h \Gamma_\mu h^\dag +h \partial_\mu h^\dag.
\eea

The transformation properties of the matter fields are not unique. It is, however, convenient to construct
the triplets for the $B$ and $D$ mesons such that they transform under $g_L\times g_R$ as
\bea
B\mapsto Bh^\dag,\qquad D\mapsto Dh^\dag,
\eea
where we have introduced the notation
\bea
B = ( B^-, B^0, B_s^0 ),\quad \text{and} \quad 
D = ( D^0, D^+, D_s^+ ).
\eea
In particular, the quantity $\Gamma_\mu$ transforms precisely as a gauge SU(3)$_V$ transformation. With
it we can construct a covariant derivative for a matter fields, e.g.,
\bea
\nabla_\mu D^\dag \al = \al \partial_\mu D^\dag +\Gamma_\mu D^\dag .
\eea

To construct the effective Lagrangian, one has to specify the power counting rules. At low energies, the momenta
as well as the masses of Goldstone bosons are counted as $\mathcal{O}(p)$~\cite{Weinberg:1978kz}. However,
the nonvanishing mass of the matter field in the chiral limit introduces new energy scales (here $M_D$ and $M_B$)
which are larger than  the hard chiral scale $\Lambda_\chi \sim 4\pi F_0$. Therefore, the temporal
component of the momenta of $D$ and $B$ mesons should be counted as $\mathcal{O}(1)$, see, e.g.,
Refs.~\cite{Gasser:1984gg,Guo:2008gp,Du:2016xbh}. Yet, the small three-momenta of the $D$- and $B$-mesons
are counted as $\mathcal{O}(p)$. To incorporate the explicit chiral breaking due to the nonvanishing light
quark masses, a spurion $\chi_\pm$ is introduced as 
\bea
\chi_\pm = u^\dag \chi u^\dag \pm u\chi^\dag u,
\eea
with
\bea
\chi = 2B_0  ~ \text{diag}(m_u,m_d, m_s),
\eea
where $B_0=|\langle 0|\bar{q}q|0\rangle|/F_0^2$ is a constant related to the quark condensate.
The surion transforms under
chiral symmetry as
\bea\label{eq:pc}
\chi_\pm \mapsto h\chi_\pm h^\dag.
\eea
The power counting rules for the building blocks of the effective Lagrangian are 
\bea
\nabla_\mu D^\dag \sim	\mathcal{O}(1), \quad 
\nabla_\mu B^\dag \sim \mathcal{O}(1), \quad
u_\mu \sim \mathcal{O}(p), \quad
\chi_\pm \sim \mathcal{O}(p^2).
\eea
With the above transformation properties and the power counting rules, one can construct the effective
chiral Lagrangian for $D$-$\phi$ interactions, e.g. Refs.~\cite{Hofmann:2003je,Guo:2008gp,Guo:2009ct,Liu:2012zya,Yao:2015qia,Du:2016ntw}.

At low energies, for the processes with $\Delta b=1$ and $\Delta c=1$ the 
interaction can be described by the effective weak 
Hamiltonian $H_\text{eff}$ which at LO has the form 
\bea\label{eq:heff}
H_\text{eff} = \frac{G_F}{\sqrt{2}}V_{cb}^\ast V_{ud}^{} \big( C_1 \mathcal{O}^d_1 
+C_2 \mathcal{O}^d_2 \big) + (b\to s) +\text{h.c.},
\eea
with $G_F$ the Fermi constant, $V_{ij}$ the Cabibbo--Kobayash--Maskawa (CKM) matrix elements, and $C_i$ the scale-dependent Wilson 
coefficients. Here, the tree-level operators read 
\bea
\mathcal{O}^d_1 \al = \al (\bar{c}_a b_b)_L(\bar{d}_bu_a)_L, \nonumber\\
\mathcal{O}^d_2 \al = \al (\bar{c}_a b_a)_L(\bar{d}_bu_b)_L, 
\eea
with the subscript $a$ and $b$ the color indices. The subscript $L$ indicates that 
only the left-hand components of the quarks are involved. 
Note that here the color space is irrelevant for our discussion, thus we simply 
drop the subscripts of $C_i$ and $\mathcal{O}_i$ hereafter.  
One can make the effective Hamiltonian fully chirally invariant by introducing 
a spurion $H_i^j$ transforming as~\cite{Bijnens:2009yr}
\bea
H_i^j \mapsto H_{i^\prime}^{j^\prime} (g_L)_i^{i^\prime} (g_L^\dag)_{j^\prime}^j.
\eea
Then the new Hamiltonian
\bea
H_\text{eff}^\prime = \frac{G_F}{\sqrt{2}}V_{cb}^\ast V_{ud}^{} H_i^j C(\bar{c} b)_L (\bar{q}_L^i q_{Lj})
\eea
is chirally invariant. For Eq.~\eqref{eq:heff}, the spurion $H_i^j$ 
(the lower index labels rows and the upper one labels columns)
corresponds to 
\bea
H = \begin{pmatrix}
0 & 0 & 0 \\
1 & 0 & 0 \\
V_{us}/V_{ud} & 0 & 0
\end{pmatrix}~ .
\eea
Here, $V_{us}/V_{ud}$ is nothing but $-\sin \theta_1$ (to be written as $-s_1$ for simplicity) with $\theta_1$ the Cabbibo angle. Then the
component $H_2^1$ describes the Cabibbo-allowed decays and $H_{3}^1$ 
the Cabibbo-suppressed ones. In the matrix form, $H$ transforms under chiral symmetry as
\bea
H \mapsto g_L^{} H g_L^\dag.
\eea
It is more convenient to introduce a homogeneously transforming suprion as
\bea
t=u H u^\dag.
\eea
With those ingredients, we are prepared to construct the effective Lagrangian describing the three-body nonleptonic decays of $B$ mesons to $D$ mesons 
and two light pseudoscalars. We are interested in the region of the 
invariant mass of a pair of the $D$ and one pseudoscalar not far from their 
threshold, such that this light pseudoscalar can be safely treated as a soft Goldstone 
boson, while the other one moves fast and can be treated as a matter field 
rather than a Goldstone boson. The fast moving pseudoscalar is realized linearly in a matrix form $M$ transforming as 
\bea
M \mapsto h M h^\dag ,
\eea
and it has the same form as $\phi$, i.e.
\bea
M =\begin{pmatrix}
\frac{1}{\sqrt{2}}\pi^0+\frac{1}{\sqrt{6}}\eta  & \pi^+ & K^+  \\
\pi^- & -\frac{1}{\sqrt{2}}\pi^0 +\frac{1}{\sqrt{6}}\eta & K^0 \\
K^- & \bar{K}^0 & -\frac{2}{\sqrt{6}}\eta
\end{pmatrix}.
\eea
Consequently, utilizing  the power counting in Eq.~\eqref{eq:pc}, 
chiral symmetry implies that the 
effective Lagrangian at $\mathcal{O}(p)$ has the form of~\cite{Savage:1989ub,Du:2017zvv}
\bea\label{lag:eff}
\lag_\text{eff} \al =\al B \Big[ c_1(u_\mu t M+Mtu_\mu ) +c_2 (u_\mu M + M u_\mu)t 
+c_3 t(u_\mu M + M u_\mu )  \nonumber \\
\al \al+ c_4 (u_\mu \langle Mt\rangle +M \langle u_\mu t\rangle 
) + c_5 t \langle M u_\mu \rangle +c_6 \langle (Mu_\mu +u_\mu M )t\rangle 
\Big] \nabla^\mu  D^\dag \nonumber \\
\al \al + B\Big[ d_1 (u_\mu tM - Mtu_\mu ) +d_2 (u_\mu M - M u_\mu ) t 
+d_3 t(u_\mu M - M u_\mu ) \nonumber \\
\al \al +d_4 (u_\mu \langle Mt \rangle - M\langle u_\mu t\rangle 
)+ d_6 \langle (Mu_\mu - u_\mu M )t\rangle \Big] \nabla^\mu D^\dag,
\eea
where the $c_i$ and $d_i$ are LECs, and $\langle\dots\rangle$ denotes the traces
in the SU(3) light-flavor space. Note that the momentum operator $\nabla_\mu$ in Eq.~\eqref{lag:eff} is chosen to act on the charmed meson field $D$. It could be on $B$ (or $M$) independently as
well. However, in our case, $M_B\gg M_D+2M_\pi$ and $M_D\gg M_\pi$ imply that
they would produce the same structures up to the LO and we can combine them by redefining the
LECs in the heavy meson limit. This effective Lagrangian considers both the chiral symmetry and
flavor SU(3) constraints (the latter  has been considered in Ref.~\cite{Savage:1989ub}). Finally, we divide the 
Lagrangaian~\eqref{lag:eff} into two groups which are symmetric and antisymmetric in 
the two light pseudoscalars, which correspond to the cases where the relative
orbital angular momentum of the light pseudoscalars pair is even and odd, 
respectively~\cite{Savage:1989ub}.

\section{$B$-meson three-body decay amplitudes}\label{sec:amp}

\begin{figure}[tb]
\begin{center}
\includegraphics[width=0.7\textwidth]{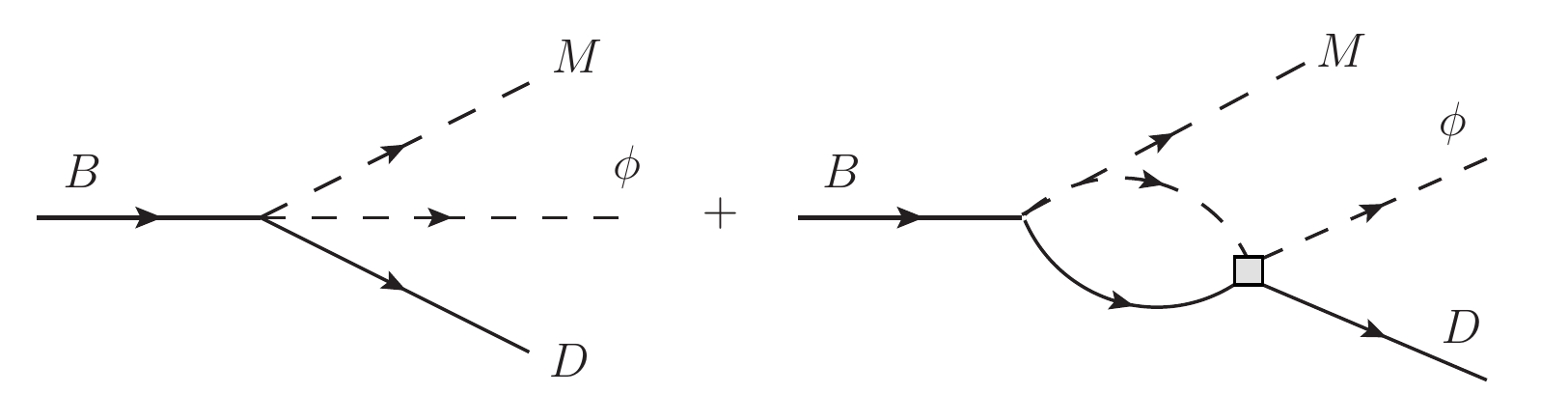}
\end{center}
\caption{Illustrative graphs for the decay $B \to D\phi M$. The square denotes 
the FSI between the $D$ meson and the Goldstone boson. In the loop, all coupled 
channels connecting the initial to the final states contribute.}\label{fig:feyndiag}
\end{figure}

The nonleptonic $B$-meson three-body decays $B\to D\phi M$, where $M$ denotes the fast moving light pesudoscalar
and $\phi$ denotes the soft one as in the last section, provide access to the $D$-$\phi$ interaction
via FSI.\footnote{We are only interested in the energy region where the $D$ and   one of the light pseudoscalars have an invariant mass not far from their threshold.} From now on, we write the decay product in the
ordering that the charm meson is followed by the soft and hard light pseudoscalar subsequently.
The FSI between the $D$ and the hard $M$ and that between
$M$ and $\phi$, which are not expected to produce any nontrivial structure sensitive to the energy variation,
are encoded into an extra complex factor for simplicity. The Feynman diagrams for the
decay $B\to D\phi M$ are shown in Fig.~\ref{fig:feyndiag}, where the square denotes the FSI between $D$ and $\phi$. 

The decay amplitude projected into the $D\phi$ channel at low energies can be decomposed into $S$-,
$P$- and $D$-waves (corresponding to the
orbital angular momentum of the $D\phi$ pair with $L=0$, $L=1$, and $L=2$, respectively),\footnote{Here,
we neglect the partial waves with $L \geq 3$, which is perfectly fine in the energy region of interest as shown in the experimental data, see e.g. Refs.~\cite{Aaij:2016fma,Aaij:2014baa,Aaij:2015vea,Aaij:2015kqa,Aaij:2015sqa}.}
\bea\label{eq:decayamplitudes}
\mathcal{A}(B\to D \phi M) = \mathcal{A}_0(s) +\sqrt{3}\mathcal{A}_1(s) 
P_1(z) +\sqrt{5}\mathcal{A}_2(s)P_2(z),
\eea
where $\mathcal{A}_{0,1,2}(s)$ correspond to the partial wave decomposed amplitudes for $D\phi$ in the $S$-, $P$-
and $D$-waves, respectively, and $P_L(z)$ are the Legendre polynomials with $z$ the cosine of the helicity
angle of the $D\phi$ system, i.e. the angle between the moving directions of the $\phi$ and the $M$ in the $D\phi$ rest frame.
For the $P$- and $D$-waves, the resonances are relatively narrow and thus it is reasonable to parameterize them
by  BW amplitudes. For the $S$-wave, we calculate the diagrams in Fig.~\ref{fig:feyndiag} with
the effective Lagrangian~\eqref{lag:eff} with the FSI provided by the $D\phi$ interaction determined in Ref.~\cite{Liu:2012zya} with the coupled-channel effects taken into account. 

For the decay $B^-\to D^+\pi^-\pi^-$, the relative orbital angular momenta of the two light mesons could
only be even, which correspond to the first term of the Lagrangian in Eq.~\eqref{lag:eff}, i.e. the
terms with the LECs $c_i$. With the Lagrangian in Eq.~\eqref{lag:eff}, the production vertices for the possible
intermediate states $D^0\pi^0$, $D^+\pi^-$, $D^0\eta$ and $D_s^+K^-$ are listed in Table~\ref{tab:DpipiS}.
\begin{table}[tbh]
\caption{Production vertices for the possible intermediate states contributing to $B^-\to D^+\pi^-\pi^-$.
Here, $p_{D_{(s)}}$ and $p_\phi$ denote the four-momenta of the charmed meson and the Goldstone boson, respectively.}
\label{tab:DpipiS}
\bea
\begin{array}{c|c}
\hline\hline
\text{Process} & \text{Production amplitude} \\
\hline
B^- \to D^0\pi^0 \pi^- & \dfrac{1}{F_0} (c_1+c_4)p_D\cdot p_\pi \\
B^- \to D^0\eta\pi^- & \dfrac{1}{\sqrt{3}F_0}(c_1+c_4+2c_2+2c_6)p_D\cdot p_\eta  \\
B^- \to D^+\pi^-\pi^- & \dfrac{2\sqrt{2}}{F_0} (c_1+c_4)p_D \cdot p_\pi \\
B^- \to D_s^+ K^-\pi^- & \dfrac{\sqrt{2}}{F_0} (c_1+c_4)p_{D_s} \cdot p_K  \\
\hline\hline
\end{array}\nonumber
\eea
\end{table}

In the heavy quark limit, $p_D\cdot p_\phi = M_D E_\phi$, with 
$E_\phi$ the energy of $\phi$ in the rest frame of $D\phi$. For convenience, we introduce
two parameters $A$ and $B$~\cite{Du:2017zvv}
\bea\label{eq:defAB}
A\al =\al \frac{\sqrt{2}}{F_0} (c_1+c_4)M_D  , \nonumber \\
B\al =\al \frac{2\sqrt{2}}{3F_0}  (c_2+c_6) M_D  .
\eea
Taking the FSI into account, the $S$-wave decay amplitude for $B^-\to D^+\pi^-\pi^-$ reads 
\bea
\mathcal{A}_0(s) \al = \al 2 A E_\pi + 2 A E_\pi G_{D\pi}(s) T_{D^+\pi^- \to D^+\pi^-}(s)  \nonumber \\
\al \al + \frac{A}{\sqrt{2}}E_\pi G_{D\pi}(s) T_{D^0\pi^0 \to D^+\pi^-} (s)
\nonumber \\
\al \al + \frac{A+3B}{\sqrt{6}}E_\eta G_{D\eta}(s) T_{D^0\eta\to D^+\pi^-} (s)
\nonumber \\
\al \al + A E_K G_{D_s\bar K}(s) T_{D_s^+K^-\to D^+\pi^-}(s),
\eea
where $s$ is the center-of-mass energy squared of the $D\phi$ system, $G_{D\phi}(s)$ is the loop function (the ``fundamental bubble'') depicted in Fig.~\ref{fig:feyndiag} 
coupling to  the channel $D\phi$. Unitarity relates the imaginary part of the loop function $G_{D\phi}(s)$
with the $D\phi$ two-body phase factor $\rho(s)=q_\phi/(8\pi\sqrt{s})\theta(s-(M_D+M_\phi)^2)$, with $q_\phi$ magnitude of the $\phi$ three-momentum in the $D\phi$ center-of-mass frame, by $\text{Im}\,G_{D\phi}(s)=-\rho(s)$, which allows us to represent 
$G_{D\phi}(s)$ via a once-subtracted dispersion relation 
\bea\label{eq:G}
G_{D\phi}(s)= \al \dfrac{1}{16\pi^2} \al \Big\{ a_A +\log\frac{M_D^2}{\mu^2} + \frac{M_\phi^2-M_D^2+s}{2s}
\log \frac{M_\phi^2}{M_D^2} \nonumber \\
\al \al + \frac{\sigma}{2s}\big[ \log (s-M_D^2+M_\phi^2+\sigma) -\log (-s+M_D^2-M_\phi^2
+\sigma) \nonumber \\
\al \al  +\log(s +M_D^2-M_\phi^2+\sigma) -\log(-s-M_D^2+M_\phi^2+\sigma )\big] \Big\},
\eea
with $a_A$ a scale-dependent subtraction constant, 
$\sigma=\{[s-(M_D+M_\phi)^2][s-(M_D-M_\phi)^2]\}^{1/2}$ and $\mu$ the scale of 
dimensional regularization. The subtraction $a_A$ is related to the renormalization of the interaction
vertices and varies for different processes. A change of $\mu$ can be absorbed into a corresponding change of $a_A$, and we will take $\mu=1$~GeV  as in Ref.~\cite{Liu:2012zya}. 

The amplitudes for the FSI can be expressed in the isospin basis. While
$D^+\pi^-$ can be decomposed into isospin $I=1/2$ and $3/2$ systems,  $D^+\eta$ and
$D_s^+ K^-$ can only form $I=1/2$. The relations between 
the isospin basis and physical particle basis are given by~\cite{Guo:2009ct,Yao:2015qia}\footnote{We use the phase convention such that $|\pi^+\rangle=-|1,+1\rangle$,
$\bar{K}^0\rangle=-|\frac{1}{2},+\frac12\rangle$, and $|D^+\rangle=-|\frac{1}{2},+\frac12 \rangle$,
where the two numbers in $|\dots\rangle$ on the right hand side are isospin $I$ and the third component $I_3$.}
\bea
T_{D^0\pi^0\to D^+\pi^-} \al = \al -\frac{\sqrt{2}}{3}T^{3/2}
_{D\pi\to D\pi} +\frac{\sqrt{2}}{3}T_{D\pi\to D\pi}^{1/2},
\nonumber \\
T_{D^0\eta^{~} \to D^+\pi^-} \al = \al \sqrt{\frac{2}{3}}
T^{1/2}_{D\eta \to D\pi}, \nonumber \\
T_{D^+\pi^-\to D^+\pi^-} \al = \al \frac{1}{3}T^{3/2}
_{D\pi \to D\pi} + \frac{2}{3}T^{1/2}_{D\pi \to D\pi}, 
\nonumber \\
T_{D_s^+K^- \to D^+\pi^-} \al = \al \sqrt{\frac{2}{3}}
T_{D_s\bar K \to D\pi}^{1/2}, 
\eea
where the superscripts indicate the total isospin $I$. The amplitudes in the isospin 
basis can be found in Refs.~\cite{Guo:2009ct,Liu:2012zya,Yao:2015qia}. As a 
result, we get the $S$-wave decay amplitude for the process $B^-\to D^+\pi^-\pi^-$~\cite{Du:2017zvv}
\bea\label{eq:swave}
\mathcal{A}_0(s) \al = \al A E_\pi \Big[ 2+ G_{D\pi}(s) \Big( \frac{5}{3}
T_{11}^{1/2}(s) +\frac{1}{3}T^{3/2}(s) \Big) \Big] 
+\frac{1}{3}(A+3B)E_\eta G_{D\eta}(s) T_{21}^{1/2}(s) \nonumber \\
\al \al +\sqrt{\frac{2}{3}}AE_KG_{D_s\bar K}(s) 
T^{1/2}_{31}(s).
\eea
Here, we write the scattering amplitudes in the matrix form $T_{ij}^I(s)$ with the
total isospin $I$, where $i$, $j$ are channel indices with 1, 2 and 3 referring 
to $D\pi$, $D\eta$ and $D_s\bar K$, respectively. Note that only two LECs 
($A$ and $B$) appear in Eq.~\eqref{eq:swave}. 

The above amplitude can also be obtained in a different way~\cite{Oset:2016lyh,Albaladejo:2016hae} without constructing the  effective Lagrangian. 
At the quark level, the decay $B^-\to D^+\pi^-\pi^-$ is mediated by the weak process 
$b\bar{u} \to (W^-c)\bar{u} \to (d\bar{u}c)\bar{u}$. 
In order to obtain a three-meson final state, an extra $q\bar{q}$ 
pair must be created from the QCD vacuum. The quantum numbers of vacuum imply that 
the extra $q\bar{q}$ pair must be a SU(3) flavor singlet, $u\bar{u}+d\bar{d}+s\bar{s}$. Next 
we hadronize the $c\bar{u}d\bar{u}(u\bar{u}+d\bar{d}+s\bar{s})$ combination in terms 
of three pseudoscalar mesons. To this end, we introduce a $q\bar{q}$ matrix 
$\Phi$ \cite{Albaladejo:2016hae},
\bea
\Phi = q \bar{q} = 
\begin{pmatrix}
u \\ d \\ s \\ c
\end{pmatrix}
\begin{pmatrix}
\bar{u} & \bar{d} & \bar{s} & \bar{c}
\end{pmatrix}
=
\begin{pmatrix}
u\bar{u} & u\bar{d} & u\bar{s} & u\bar{c} \\
d\bar{u} & d\bar{d} & d\bar{s} & d\bar{c} \\
s\bar{u} & s\bar{d} & s\bar{s} & s\bar{c} \\
c\bar{u} & c\bar{d} & c\bar{s} & c\bar{c} 
\end{pmatrix}
,
\eea
which fulfills 
\bea
\Phi \Phi = (q\bar{q})(q\bar{q}) = q(\bar{q}q) \bar{q}=(\bar{u}u+\bar{d}d
+\bar{s}s+\bar{c}c)\Phi ,
\eea
where the singlet $(\bar{q}q)=(\bar{u}u+\bar{d}d +\bar{s}s+\bar{c}c)$ corresponds 
to a $\bar{q}q$ pair creation from the QCD vacuum. 
\begin{figure}[tbh]
\begin{center}
\includegraphics[width=0.7\textwidth]{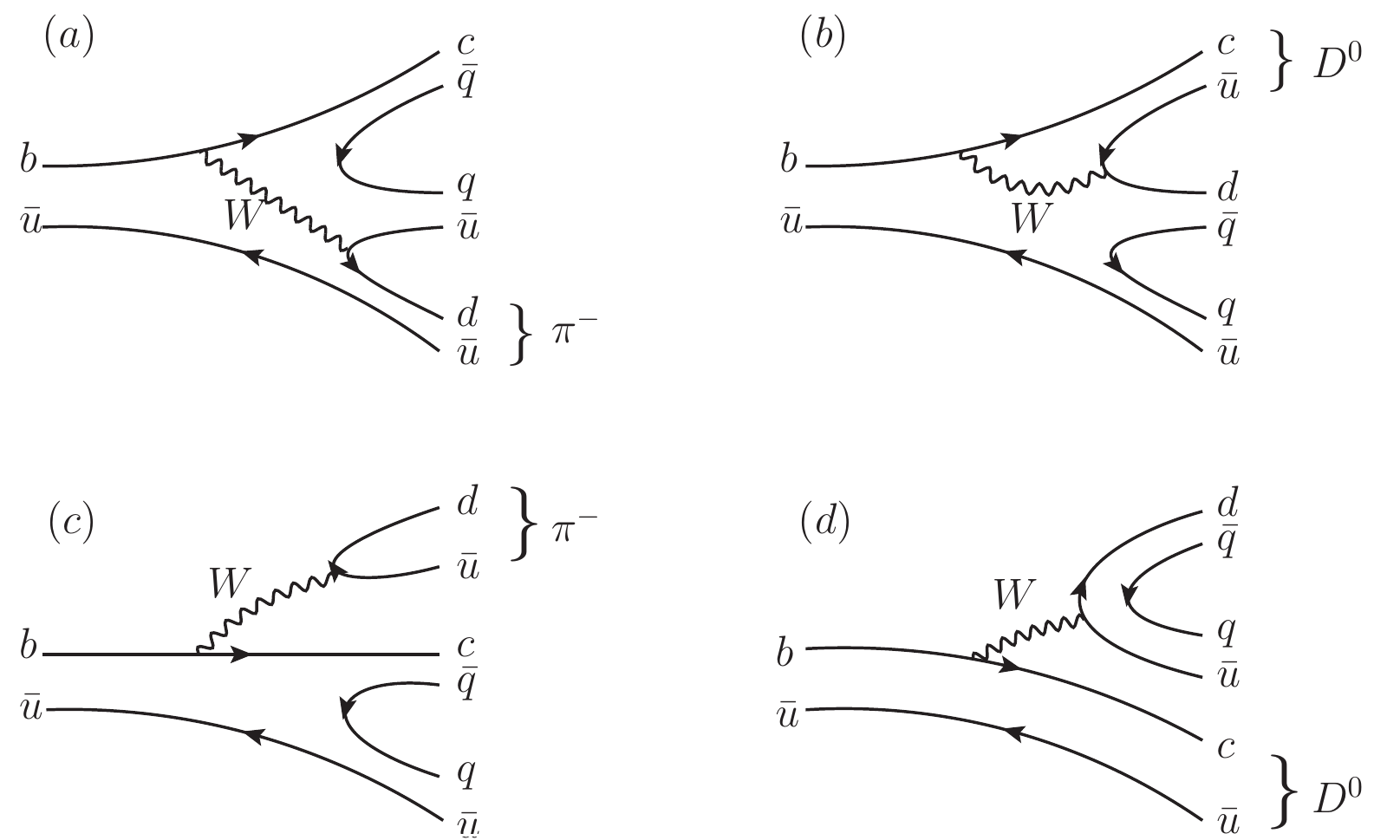}
\end{center}
\caption{Schematic representation at the quark level for the decay of the $B^-$ meson into  $D^0\pi^0\pi^-/
  D^+\pi^-\pi^-/ D^0\eta\pi^-/ D_s^+K^-\pi^-$.}\label{fig:decayquark}
\end{figure}
In terms of pseudoscalar mesons,  the matrix $\Phi$ has the form of 
\bea
\Phi = \begin{pmatrix}
\frac{1}{\sqrt{2}}\pi^0 +\frac{1}{\sqrt{6}}\eta & \pi^+ & K^+ & \bar{D}^0 \\
\pi^- & -\frac{1}{\sqrt{2}}\pi^0 +\frac{1}{\sqrt{6}}\eta & K^0 & D^- \\
K^- & \bar{K}^0 & -\frac{2}{\sqrt{6}}\eta & D_s^- \\
D^0 & D^+ & D_s^+ & \eta_c
\end{pmatrix}
~.
\eea
The hadronization of $c\bar{u}d\bar{u}(q\bar{q})$ could proceed through two 
distinct ways, i.e., $[d\bar{u}] (\Phi \Phi)_{41}$
and $[c\bar{u}](\Phi \Phi)_{21}$, which correspond to the diagrams $(a,c)$, and $(b,d)$, 
respectively, in Fig.~\ref{fig:decayquark}. These two sets of hadronization 
give two independent structures as
\bea
[d\bar{u}](\Phi\Phi)_{41} \al = \al \pi^- \left( \frac{1}{\sqrt{2}}D^0\pi^0 
+\frac{1}{\sqrt{6}}D^0\eta + D^+\pi^- +D_s^+K^- \right) +\cdots ,
\eea
and
\bea
[c\bar{u}](\Phi\Phi)_{21} \al = \al D^0 \left( \sqrt{\frac{2}{3}}  \eta\pi^- \right)+\cdots , \qquad \qquad\qquad\qquad\qquad\qquad~
\eea
respectively. We assume that the production amplitudes for the diagrams in Fig.~\ref{fig:decayquark} are $a$, $b$, $c$ and 
$d$, respectively. Then the amplitudes for  the processes in Table~\ref{tab:DpipiS} are proportional to the following factors:
\bea\label{eq:quarkamp}
\frac{a+c}{\sqrt{2}},\quad \frac{(a+c)+2(b+d)}{\sqrt{6}}, \quad
2(a+c),\quad \text{and}\quad (a+c),
\eea
respectively. Compared with the expressions in Table~\ref{tab:DpipiS}, we find $A=a+c$ and 
$B=2(b+d)/3$. Then the SU(3) flavor structure for the amplitudes of the four processes in Table~\ref{tab:DpipiS} reads
\bea
\frac{1}{\sqrt{2}}A : \left(\frac{1}{\sqrt{6}}A+\sqrt{\frac{3}{2}}B\right):2A:A\,.
\eea
Expressed in the isospin basis,
the amplitudes for the production of the $D\pi(I=1/2)$, $D\pi(I=3/2)$, $D\eta(I=1/2)$ and $D_s\bar K(I=1/2)$ channels have the ratios
\bea
\frac{5}{\sqrt{6}}A:\left(-\frac{1}{\sqrt{3}}A\right):\left(\frac{1}{\sqrt{6}}A+\sqrt{\frac{3}{2}}B\right):A\,.
\eea
Since the $D\pi$ systems with $I=1/2$ and $3/2$ have both the $D^0\pi^0$ and the $D^+\pi^-$ components, we need to project out
the $D^+\pi^-$ component and obtain the ratio for $D^+\pi^-(I=1/2)$ and $D^+\pi^-(I=3/2)$. As a result,
trivial overall factors $\sqrt{\frac{2}{3}}$ and $-\sqrt{\frac{1}{3}}$ are needed for $I=1/2$ and $3/2$,
respectively. Taking into account the chiral symmetry implying that the amplitudes are proportional to the energy of the Goldstone bosons and
the FSI,  we reobtain the amplitude in Eq.~\eqref{eq:swave}. 

With the Lagrangian in Eq.~\eqref{lag:eff}, one can calculate the production vertices responsible for other processes
$B_s^0\to \bar{D}^0K^-\pi^+$, $B^0\to\bar{D}^0\pi^-\pi^+$, $B^-\to D^+\pi^-K^-$, and $B^0\to\bar{D}^0\pi^-K^+$
as well. For these reactions experimental data are available from the LHCb experiment~\cite{Aaij:2014baa,Aaij:2015vea,Aaij:2015kqa,Aaij:2015sqa}. The weak production vertices needed for
those decays are listed in Table~\ref{tab:amps}. Note that in Ref.~\cite{Du:2017zvv}, only the
terms corresponding to an even relative angular momentum of the two light mesons are considered, i.e. the first
term in the Lagrangian~\eqref{lag:eff}. It is appropriate for the decay $B^-\to D^+\pi^-\pi^-$
and numerically sufficient to describe the decay $B_s^0\to \bar{D}^0K^-\pi^+$. We find that the inclusion of the
terms for the two light mesons with an odd relative angular momentum does not improve the fit quality significantly,
which means that the experimental data used in Ref.~\cite{Du:2017zvv} are not sufficient to disentangle the
contributions from the two terms. In this paper, more experimental data, i.e., $B^0\to\bar{D}^0\pi^-\pi^+$,
$B^-\to D^+\pi^-K^-$, and $B^0\to\bar{D}^0\pi^-K^+$, are considered and a discrimination of both contributions is possible. 
\begin{table}[tbh]
\caption{Weak amplitudes contributing to the decays $B_s^0\to \bar{D}^0K^-\pi^+$, $B^0\to\bar{D}^0\pi^-\pi^+$,
$B^-\to D^+\pi^-K^-$, and $B^0\to\bar{D}^0\pi^-K^+$ through coupled-channel effects. 
}
\label{tab:amps}
\bea
\begin{array}{l|c}
\hline\hline
\text{Processes} & \text{Weak production vertices} \\
\hline
B_s^0\to \bar{D}^0K^-\pi^+ & \dfrac{\sqrt{2}M_D}{F_0}E_K\Big( (c_2+c_4) + (d_2+d_4)\Big)  \\
B_s^0\to D^-\bar{K}^0 \pi^+ & \dfrac{\sqrt{2}M_D}{F_0}E_K\Big((c_1+c_4) +(d_1+d_4)
\Big) \\
B_s^0\to \bar{D}_s\eta\pi^+ & \dfrac{2M_{D_s}}{\sqrt{3}F_0}E_\eta\Big( (c_6-c_4)-d_4  \Big) \\
B_s^0\to \bar{D}_s\pi^0\pi^+ & \dfrac{2M_{D_s}}{F_0}E_\pi d_6 \\
\hline
B^0\to \bar{D}^0\pi^-\pi^+ & \dfrac{\sqrt{2}M_D}{F_0}E_\pi\Big( (c_2+c_3+c_4+2c_5)+(d_2-d_3+d_4)\Big) \\
B^0\to D^-\eta \pi^+ & \dfrac{M_D}{\sqrt{3}F_0}E_\eta \Big( (c_1+2c_3+c_4+2c_6)+(d_1+d_4)\Big) \\
B^0\to D_s^- K^0\pi^+ & \dfrac{\sqrt{2}M_{D_s}}{F_0}E_K\Big( (c_3+c_4) -(d_3-d_4) \Big)\\
B^0\to  D^- \pi^0\pi^+ & -\dfrac{M_D}{F_0}E_\pi\Big( (c_1+c_4) +(d_1-2d_3+d_4-2d_6)\Big) \\
\hline
B^0\to \bar{D}^0\pi^-K^+ & -s_1\dfrac{\sqrt{2}M_D}{F_0}E_\pi\Big( (c_2+c_4)+(d_2+d_4)\Big)\\
B^0\to D^-\pi^0K^+ & -s_1\dfrac{M_D}{F_0}E_\pi\Big( -(c_4-c_6) -(d_4-d_6) \Big) \\
B^0\to D^-\eta K^+ & -s_1\dfrac{M_D}{\sqrt{3}F_0}E_\eta \Big( (c_4-c_6)+(d_4+3d_6)\Big) \\
B^0\to D_s^-K^0 K^+ & -s_1\dfrac{\sqrt{2}M_{D_s}}{F_0}E_K \Big( (c_1+c_4)+(d_1+d_4) \Big) \\
\hline
B^-\to D^0\pi^0 K^- & -s_1 \dfrac{M_D}{F_0}E_\pi \Big( (c_1+c_4+c_2+c_6) -(d_1-d_2-d_4-d_6)\Big)\\
B^-\to D^0\eta K^- & -s_1 \dfrac{M_D}{\sqrt{3}F_0}E_\eta \Big( (c_1+c_4-c_2-c_6) -(d_1-3d_2-d_4-3d_6) \Big) \\
B^-\to D^+\pi^- K^- & -s_1\dfrac{\sqrt{2}M_D}{F_0}E_\pi \Big( (c_1+c_4) -(d_1-d_4) \Big) \\
B^-\to D_s^+K^- K^+ & -s_1\dfrac{2\sqrt{2}M_{D_s}}{F_0} E_K(c_1+c_4 )\\
\hline\hline
\end{array}\nonumber
\eea
\end{table}

Although these decays have three hadrons in the final states, we expect that, at least in the
low-energy tails of the invariant mass of the $D\phi$ subsystems, the effects from the crossed-channel FSIs (interactions between 
the soft and hard light mesons and those between the $D$ meson and
the hard pseudoscalar meson) do not produce any nontrivial structure in the $D\phi$ distributions, which is supported by the analyses in Refs.~\cite{Aaij:2014baa,Aaij:2015vea,Aaij:2015kqa,Aaij:2015sqa}. Therefore, for a description of the low-energy $D\phi$ systems, we do not need to account for the full three-body FSIs. The effects from the crossed-channel FSIs can thus be simply
encoded into an extra undetermined complex factor for each partial wave as it is done in the isobar model.
Following the procedures for the $B^-\to D^+\pi^-\pi^-$ above, one can obtain the $S$-wave amplitudes as\footnote{Hereafter,
we neglect the overall factor $\frac{\sqrt{2}M_D}{F_0}$ for the $S$-wave amplitudes by absorbing it into the LECs.}
\bea\label{eq:amp:1}
\mathcal{A}_0(B_s^0\to \bar{D}^0K^-\pi^+)\al = \al  (c_2+c_4+d_2+d_4)E_K +
d_6 E_\pi G_{D_s\pi}(s)T_{D_s\pi\to\bar{D}\bar{K}}^1(s) \nonumber\\
\al \al + \frac{1}{2}(c_2-c_1+d_2-d_1)E_K G_{DK}(s) T_{\bar{D}\bar{K}\to \bar{D}\bar{K}}^1(s) \nonumber\\
\al \al + \frac{1}{2}(c_1+c_2+2c_4+d_1+d_2+2d_4)E_K G_{DK}(s) T_{\bar{D}\bar{K}\to \bar{D}\bar{K}}^0(s) \nonumber\\
\al \al + \sqrt{\frac13}(c_4-c_6+d_4)E_\eta G_{D_s\eta}(s)T_{\bar{D}_s\pi\to \bar{D}\bar{K}}^0(s),
\eea
\bea\label{eq:amp:2}
\mathcal{A}_0(B^0\to \bar{D}^0\pi^-\pi^+)\al=\al (c_2+c_3+c_4+2c_5+d_2-d_3+d_4)E_K +\frac13E_\pi G_{D\pi}(s) T_{D\pi\to D\pi}^{1/2}(s)\nonumber\\
\al \al ~ \times (c_1+2c_2+2c_3+3c_4+4c_5+d_1+2d_2-4d_3+3d_4-2d_6)\nonumber\\
\al \al + \frac{1}{3} (c_2-c_1+c_3+2c_5-d_1+d_2+d_3+2d_6)E_\pi G_{D\pi}(s)T_{D\pi\to D\pi}^{3/2}(s) \nonumber\\
\al \al + \frac13 (c_1+2c_3+c_4+2c_6+d_1+d_4)E_\eta G_{D\eta}(s)T_{D\eta\to D\pi}^{1/2}(s) \nonumber\\
\al \al + \sqrt{\frac23}(c_3+c_4-d_3+d_4)E_KG_{D_s\bar K}(s)T_{D_s\bar{K}\to D\pi}^{1/2}(s),
\eea
\bea\label{eq:amp:3}
\mathcal{A}_0(B^-\to D^+\pi^- K^-) \al =\al -s_1 (c_1+c_4-d_1+d_4)E_\pi -2\sqrt{\frac23}s_1(c_1+c_4)E_KG_{D_s\bar K}(s)T^{1/2}_{D_s\bar{K}\to D\pi}(s) \nonumber\\
\al \al - s_1(3c_1+c_2+3c_4+c_6-3d_1+d_2+3d_4+d_6)E_\pi G_{D\pi}(s)T^{1/2}_{D\pi\to D\pi}(s) \nonumber \\
\al \al - s_1(c_1-c_2+c_4-c_6-d_1+d_4+3d_2+3d_6)E_\eta G_{D\eta}(s)T^{1/2}_{D\eta\to D\pi}(s) \nonumber \\ 
\al \al + \frac13 s_1(c_2+c_6+d_2+d_6) E_\pi G_{D\pi}(s)T^{3/2}_{D\pi\to D\pi}(s),\\
\label{eq:amp:4}
\mathcal{A}_0(B^0\to \bar{D}^0 \pi^- K^+) \al =\al -s_1(c_2+c_4+d_2+d_4)E_\pi  \nonumber\\
\al \al- \frac13 s_1(2c_2+3c_4-c_6+2d_2+3d_4-d_6)E_\pi G_{D\pi}(s)T_{D\pi\to D\pi}^{1/2}(s)\nonumber\\
\al \al - \frac13 s_1(c_4-c_6+d_4+3d_6)E_\eta G_{D\eta}(s)T_{D\eta \to D\pi}^{1/2}(s)\nonumber \\
\al \al- \sqrt{\frac23}s_1(c_1+c_4+d_1+d_4)E_K G_{D_s\bar K}(s)T_{D_s\bar{K}\to D\pi}^{1/2}(s) \nonumber\\
\al \al - \frac13 s_1(c_2+c_6+d_2+d_6)E_\pi G_{D\pi} T^{3/2}_{D\pi\to D\pi}(s).
\eea

Next, we show that the $S$-wave decay amplitudes above satisfy the two-body unitarity relation
\bea
\mathcal{A}-\mathcal{A}^\ast =-2iT\rho \mathcal{A}^\ast= -2i T^\dag \rho \mathcal{A}, \label{eq:Auni}
\eea
where $T$ represents the $T$-matrix for the two-body scattering and fulfills the unitarity relation\footnote{Here,
$T$ is defined via $S=1-iT$, with $S$ the $S$-matrix.}
\bea\label{eq:scatuni}
T-T^\dag = -2iT^\dag \rho T,
\eea
which can be rewritten as
\bea\label{eq:scatuni2}
(T^\dag)^{-1}-T^{-1}=-2i\rho.
\eea
The $T$-matrix in Ref.~\cite{Liu:2012zya} is given by 
\bea\label{eq:T}
T=V[1-\tilde{G}V]^{-1}=[V^{-1}-\tilde{G}]^{-1},
\eea
where $V$ is the $S$-wave projection of the $D\phi$ scattering potentials from ChPT and $\tilde{G}$ is a diagonal matrix whose nonvanishing elements are the loop functions with the same
form as the $G(s)$ function in Eq.~\eqref{eq:G}~\cite{Liu:2012zya}. It can be rewritten in a more concise form
\bea\label{eq:Tinv}
T^{-1}=V^{-1}-\tilde{G}.
\eea
Since the $\tilde{G}$ matrix elements satisfy $\text{Im}\,\tilde{G}_{ii}(s)=-\rho_{ii}(s)$, with $\rho_{ii}(s)$ the two-body phase space factor in the $i$-th channel,
and $V(s)$ is real for real $s$, it is easy to conclude that the $T$-matrix fulfills the
two-body unitarity in Eqs.~(\ref{eq:scatuni},\ref{eq:scatuni2}). 

To show that the decay amplitudes in Eq.~\eqref{eq:swave}, (\ref{eq:amp:1}-\ref{eq:amp:4}) fulfill the two-body unitarity in Eq.~\eqref{eq:Auni},
we write them in a matrix form
\bea\label{eq:Amatrix}
\mathcal{A}=P + T GP.
\eea
In components, this reads
\bea
\mathcal{A}_i =P_i +\sum_j P_j G_j T_{ji},
\eea
where $P_i$ represents the weak production vertices given in Table~\ref{tab:amps}. Making use of Eq.~\eqref{eq:Tinv},
the decay amplitudes in Eq.~\eqref{eq:Amatrix} can be rewritten as 
\bea\label{eq:Arewrite}
\mathcal{A}=(1+TG)P=T(T^{-1}+G)P=T(V^{-1}+G-\tilde{G})P.
\eea
Because of the time reversal symmetry of the strong interaction, which means that $T^\dag=T^\ast$, one obtains
the desired two-body unitarity for the decay amplitudes
\bea
-2iT^\dag \rho \mathcal{A}\al = \al -2iT^\dag \rho T (V^-2 +G-\tilde{G})P\nonumber \\
\al = \al (T-T^\ast)(V^{-1}+G-\tilde{G})P \nonumber\\
\al = \al \mathcal{A}-\mathcal{A}^\ast.
\eea

Having constructed  the $S$-wave amplitudes, the complex decay amplitudes for $P$- and $D$-waves are described
by an isobar model as coherent sums of intermediate resonant decays. This is reasonable because of the
relatively narrow widths of the resonances in $P$- and $D$-waves. The $P$- and $D$-wave amplitudes used
in Ref.~\cite{Du:2017zvv} have the same BW form as that in the LHCb analysis~\cite{Aaij:2016fma}. For the
low-energy $D^+\pi^-$ systems in processes $B^-\to D^+\pi^-\pi^-$ and $B^-\to D^+\pi^- K^-$, the $P$-
and $D$-wave resonant contributions considered here include the ground state charmed vector meson $D^{\ast }$ and an excited state $D_1^\ast(2680)$~\cite{Aaij:2016fma} in the $P$-wave, and the $D_2^\ast(2460)$ in the $D$-wave as in
Ref.~\cite{Du:2017zvv}. For these decays, the $P$- and $D$-wave amplitudes are given by
\bea\label{eq:resonance1}
\mathcal{A}_1(B^- \to D^+\pi^- M) \al = \al c_{D^\ast}F_{D^\ast}(s,z)+c_{D^\ast}^\prime F_{D_1^\ast(2680)}(s,z), \nonumber\\
\mathcal{A}_2(B^- \to D^+\pi^- M) \al = \al c_{D_2}F_{D_2}(s,z),
\eea
where the complex coefficients $c_{D^*}$, $c_{D^*}'$ and  $c_{D_2}$ 
vary for different decays and will be determined by fitting to the experimental data. $F_i(s,z)$
represents the contribution of each resonance to the corresponding decay. For the decays $B^0\to \bar{D}^0\pi^- \pi^+$
and $B^0\to \bar{D}^0\pi^- K^+$, the vector and tensor charmed mesons $D^{\ast}$ and $D^\ast_2(2460)$ are taken into
account as those in the LHCb analysis~\cite{Aaij:2015sqa,Aaij:2015kqa},
\bea\label{eq:resonance2}
\mathcal{A}_1(B^0 \to \bar{D}^0\pi^-  M) \al = \al c_{D^\ast}F_{D^\ast}(s,z), \nonumber\\
\mathcal{A}_2(B^0 \to \bar{D}^0\pi^-  M) \al = \al c_{D_2}F_{D_2}(s,z).
\eea

For the decay $B_s^0\to \bar{D}^0K^-\pi^+$, we are interested in is the low-energy region of the $\bar{D}^0K^-$. The strange partners of the ones in Eq.~\eqref{eq:resonance2}, i.e. the
$D_s^\ast$ in $P$-wave and $D_{s2}(2573)$ in $D$-wave are included as
\bea\label{eq:resonance3}
\mathcal{A}_1(B_s^0\to \bar{D}^0K^-\pi^+) \al = \al c_{D_s^\ast}F_{D_s^\ast}(s,z) , \nonumber\\
\mathcal{A}_2(B_s^0\to \bar{D}^0K^-\pi^+) \al = \al c_{D_{s2}}F_{D_{s2}}(s,z).
\eea
$F_i(s,z)$ is a function of both the  $D\phi$ invariant mass and $z$, the cosine of the $D\phi$ helicity angle.  It has the form
\bea
F^{(J)}(s,z)=R^{(J)}(s)\times X^{(J)}(|\vec{p}|r_\text{BW})\times X^{(J)}(|\vec{q}|r_\text{BW})
\times M^{(J)}(\vec{p},\vec{q}),
\eea
where $\vec{p}$ and $\vec{q}$ are the momenta of the hard light meson $M$ and the one of the resonance
daughters, respectively, both evaluated in the rest frame of the resonance. 
Here, the relativistic BW propagator is given by
\bea
R^{(J)}(m^2) \al =\al \frac{1}{(m_0^2-m^2)-im_0\Gamma^{(J)}(m)}, \nonumber\\
\Gamma^{(J)}(m)\al =\al \Gamma_0 \Big(\frac{q}{q_0}\Big)^{2J+1}\Big(\frac{m_0}{m}\Big)X^{(J)2}(qr_\text{BW}),
\eea
where $q=|\vec{q}|$, $\Gamma_0$ is the width of the static resonance, and  $q_0$ is $q$ evaluated at the resonance mass $m_0$.
The $P$- and $D$-wave Blatt--Weisskopf barrier  factors $X^{(J)}(x)$ are given by 
\bea
X^{(1)}(x)\al = \al \sqrt{\frac{1+x_0^2}{1+x^2}}, \nonumber\\
X^{(2)}(x)\al = \al \sqrt{\frac{x_0^4+3x_0^2+9}{x^4+3x^2+9}},
\eea
where $x_0$ represents the value of $x$ evaluated at $m=m_0$.
The radius of the barrier, $r_\text{BW}$, is taken to be 4.0~GeV$^{-1}$~\cite{Aaij:2014baa,Aaij:2015vea,Aaij:2015kqa,Aaij:2015sqa,Aaij:2016fma}. The angular probability distribution
terms, $M^{(J)}$, are given in the Zemach tensor formalism~\cite{Zemach:1963bc} by
\bea
M^{(1)}(\vec{p},\vec{q}) \al  = \al -2\vec{p}\cdot \vec{q}, \nonumber\\
M^{(2)}(\vec{p},\vec{q}) \al = \al \frac43 \Big( 3(\vec{p}\cdot \vec{q} )^2-(|\vec{p}||\vec{q}|)^2\Big),
\eea
which have similar structures as the Legendre polynomials referring to the angle between $\vec{p}$ and
$\vec{q}$. The virtual contributions from the resonances with masses outside the kinematically allowed
region, e.g. $D^\ast$, are described by the function $F_i(s,t)$ as well, however, with the modification that
the mass $m_0$ is replaced by an effective mass $m_0^\text{eff}$ which is given by (see, e.g., Ref.~\cite{Aaij:2016fma}) 
\bea
m_0^\text{eff} = m^\text{min}+(m^\text{max}-m^\text{min})\Bigg(1+\text{tanh}\frac{m_0-\frac{m^\text{min}
+m^\text{max}}{2}}{m^\text{max}-m^\text{min}} \Bigg),
\eea
where $m^\text{max}$ and $m^\text{min}$ represents the upper and lower limits of the kinematically allowed mass range.

\section{Numerical analysis}\label{sec:fit}

In this section, we demonstrate that the amplitudes in Eqs.~\eqref{eq:amp:1}-\eqref{eq:amp:4} can indeed describe the
data collected at the LHCb experiment in Refs.~\cite{Aaij:2014baa,Aaij:2015vea,Aaij:2015kqa,Aaij:2015sqa}.
We fit to the so-called angular moments, which contain important information about the
partial-wave phase variations. The angular moments $\langle P_L\rangle$ are obtained by weighting the
event distribution in the invariant mass by the Legendre polynomial of order $L$ with respect to $z$,
\bea
\langle P_L(s)\rangle = \int_{-1}^{1} d z \frac{d\Gamma}{d\sqrt{s} dz}P_L(z).
\eea
The angular moments are most powerful when a resonance is present only in one invariant mass combination.
The structures show up in moments up to 2$J$, where $J$ is the spin of the contributing resonance~\cite{Aaij:2014baa}. Neglecting partial waves with $L \geq 3$, the first few unnormalized moments (normalized relative to each another) are given by 
\bea\label{eq:angularm}
\langle P_0\rangle \al \propto \al |\mathcal{A}_0|^2+|\mathcal{A}_1|^2+|\mathcal{A}_2|^2 , \nonumber \\
\langle P_1 \rangle \al \propto \al \dfrac{2}{\sqrt{3}}|\mathcal{A}_0||\mathcal{A}_1 | \cos (\delta_0-\delta_1 ) +\dfrac{4}{\sqrt{15}} |\mathcal{A}_1| 
|\mathcal{A}_2| \cos (\delta_1 -\delta_2 ), \nonumber \\
\langle P_2 \rangle \al \propto \al \dfrac{2}{5}|\mathcal{A}_1|^2 
+ \dfrac{2}{7}|\mathcal{A}_2|^2 +\dfrac{2}{\sqrt{5}}|\mathcal{A}_0|
|\mathcal{A}_2| \cos (\delta_0-\delta_2 ), \nonumber \\
\langle P_3\rangle \al \propto \al \dfrac{6}{7}\sqrt{\dfrac{3}{5}}
|\mathcal{A}_1||\mathcal{A}_2|\cos (\delta_1-\delta_2) ,
\eea
where $\delta_i$ is the phase of $\mathcal{A}_i$, i.e., $\mathcal{A}_i=|\mathcal{A}_i|e^{i\delta_i}$. Instead
of $\langle P_1\rangle$ and $\langle P_3\rangle$, we analyze their linear combination  as proposed in Ref.~\cite{Du:2017zvv},
\bea\label{eq:angularm13}
\langle P_{13}\rangle = \langle P_1\rangle -\frac{14}{9}\langle P_3\rangle 
\propto \frac{2}{\sqrt{3}} |\mathcal{A}_0||\mathcal{A}_1|\cos (\delta_0-\delta_1),
\eea
which only depends on the $S$-$P$ interference up to $L=2$ and is particularly sensitive to the $S$-wave phase motion. 

Before fitting to the experimental data of the angular moments, let us investigate the $S$-wave
amplitudes in Eqs.~(\ref{eq:amp:1}-\ref{eq:amp:4}) in more detail. Although there are 11 unknown LECs in the
Lagrangian~\eqref{lag:eff}, only 10 combinations are independent.
Furthermore, to reduce the their correlations in the fit procedure, the LECs are redefined as 
\bea
A=c_1+c_4 , \quad B= \frac32 (c_2+c_6), \quad C= c_2+c_4 , \quad D= c_3+2c_5,
\quad E = c_3+c_6 , \nonumber \\
A^\prime = d_1-d_4, ~~\quad B^\prime = d_2+d_6, \quad C^\prime = d_4-d_6, \quad 
D^\prime = d_4+d_6, \quad E^\prime = d_3.\quad ~~ 
\eea
In particular, the definitions of $A$, $B$ are same to that in Ref.~\cite{Du:2017zvv}, and thus we will the ratio $B/A$
determined therein by fitting to the high quality data for the decay $B^-\to D^+\pi^-\pi^-$, in which $B/A$ and
the subtraction constant in the $G$ function are the only free parameters for the $S$-wave amplitude.\footnote{The value of $A$ cannot be determined since it serves as the normalization constant for the $S$-wave contribution as well.}
Moreover, although the subtraction constants for different channels could be different in principle, as in Ref.~\cite{Du:2017zvv},
a uniform value is taken to reduce the number of free parameters. For the $P$- and
$D$-wave amplitudes in Eqs.~(\ref{eq:resonance1}-\ref{eq:resonance3}), two real parameters, the magnitude and the
phase of the corresponding $c_D$, are needed for each resonance. Since we are not interested in the precise $P$- and $D$-wave
amplitudes, the resonance mass and width are taken as the central values reported in the LHCb analyses~\cite{Aaij:2014baa,Aaij:2016fma} and the uncertainties therein are not considered.
The used resonance parameters are listed in Table~\ref{tab:resonance}.
\begin{table}[tbh]
  \caption{The masses and widths of the $P$- and $D$-wave resonances are taken from the LHCb analyses~\cite{Aaij:2014baa,Aaij:2016fma}.}\label{tab:resonance}
\vspace{-0.5cm}
\bea
\begin{array}{lccc}
\hline
\hline
\text{Resonance} & \text{Spin} & \text{Mass (MeV)} & \text{Width (MeV)}  \\
\hline
D^\ast & 1 &  2006.98 & 2.1 \\
D_2^\ast(2460) & 2 & 2463.7 & 47.0  \\
D_1^\ast(2680) & 1 & 2681.1  & 186.7  \\
D_s^\ast & 1 &  2112.1 & 143.8 \\
D_{s2}(2573) & 2 & 2568.39 & 16.9  \\
\hline\hline
\end{array}\nonumber
\eea
\end{table}

\begin{figure*}[tb]
  \begin{center}
   \includegraphics[width=0.31\linewidth]{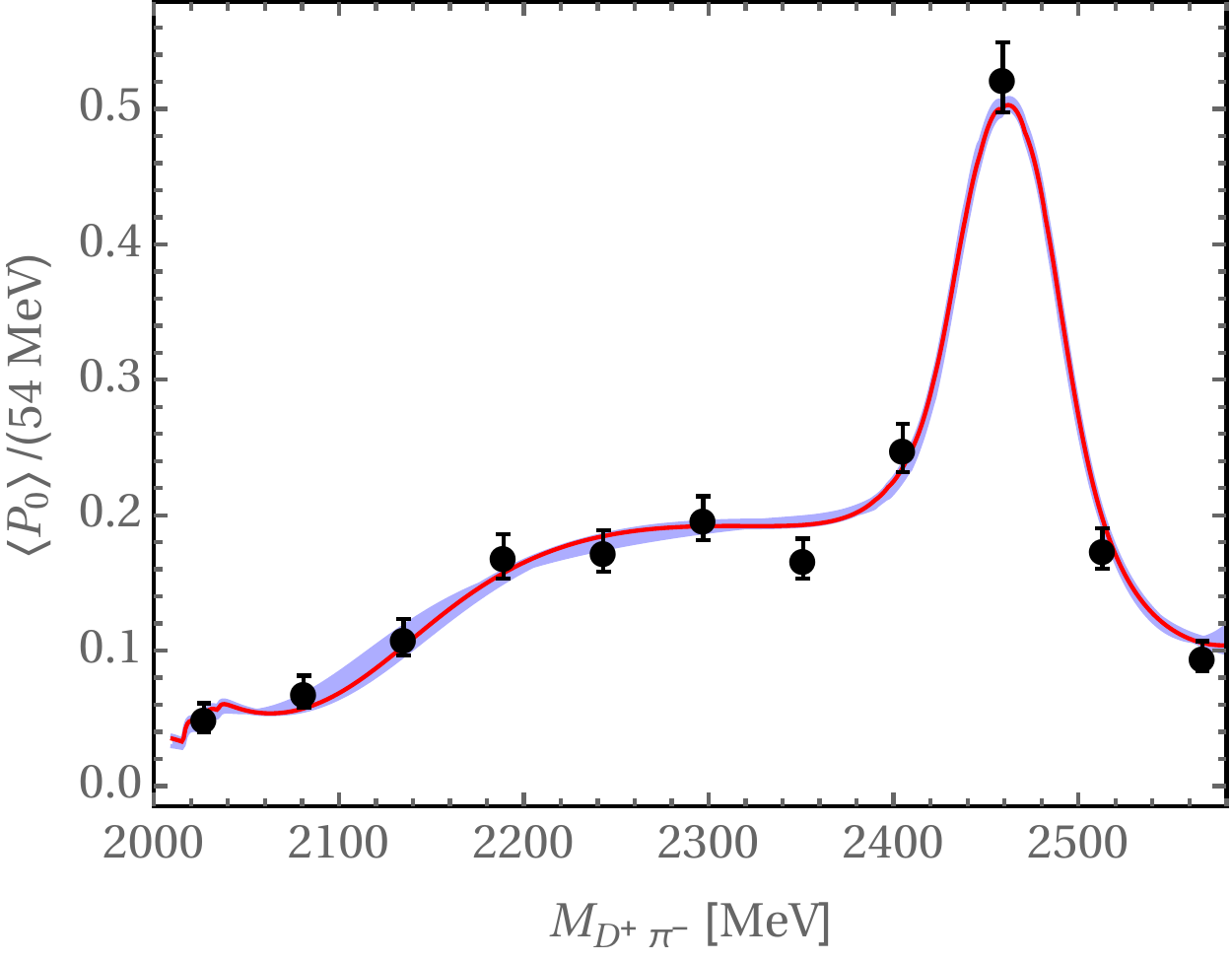} \hfill
  \includegraphics[width=0.32\linewidth]{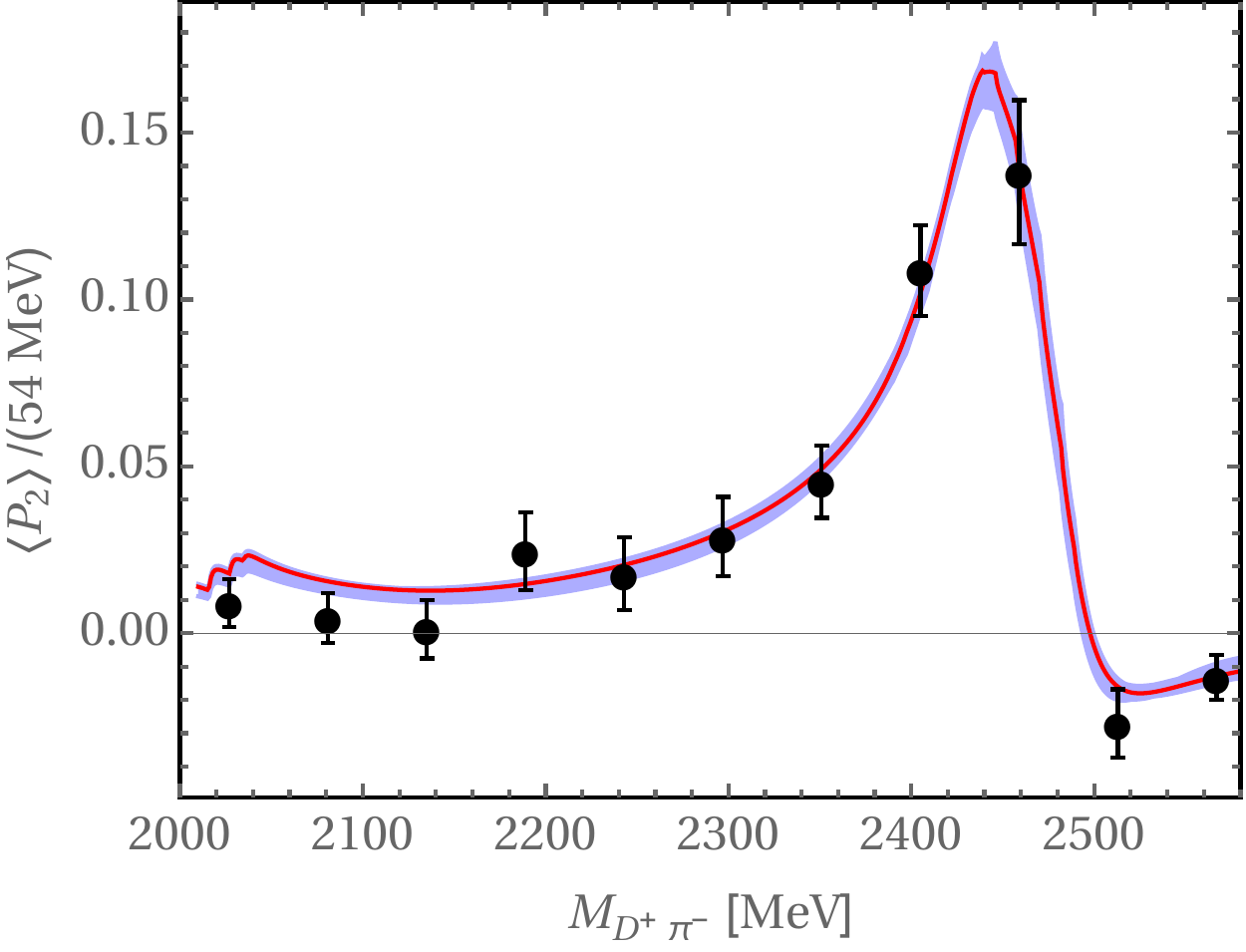} \hfill
  \includegraphics[width=0.31\linewidth]{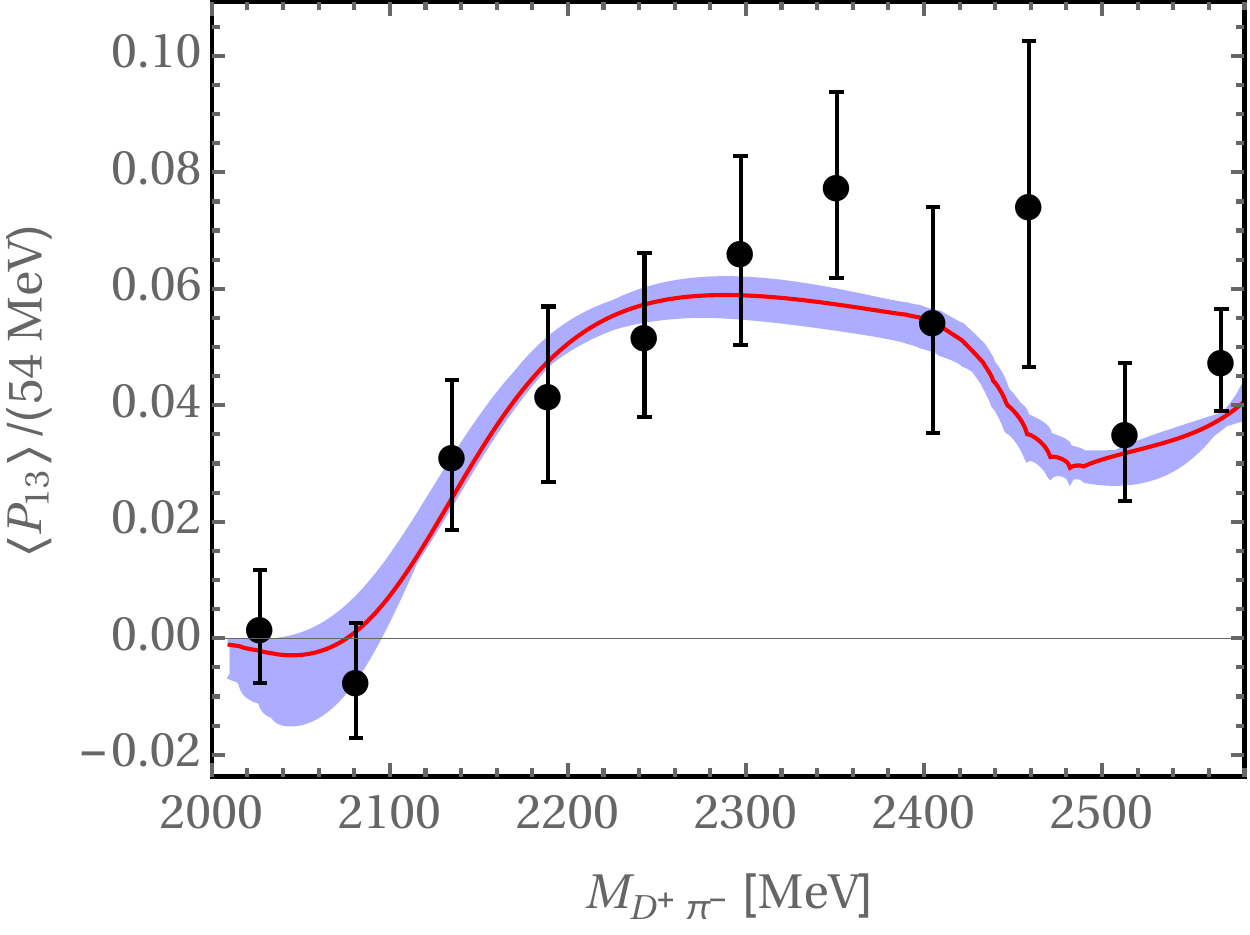}
 \end{center}
  \caption{
Fit to the LHCb data of the angular moments $\langle P_0\rangle$, $\langle P_2\rangle$ and $\langle P_{13}\rangle$
for the $B^-\to D^+\pi^- K^-$ reaction~\cite{Aaij:2015vea}. 
The largest error of $\langle P_1\rangle$ and $14\langle P_3\rangle/9$ in each bin is taken as
the error of $\langle P_{13}\rangle$. The solid lines show  
our best fit results with error bands corresponding to the one-sigma uncertainties propagated 
from the input scattering amplitudes.
\label{fig:fit1}}
\end{figure*}

\begin{figure*}[tb]
  \begin{center}
   \includegraphics[width=0.31\linewidth]{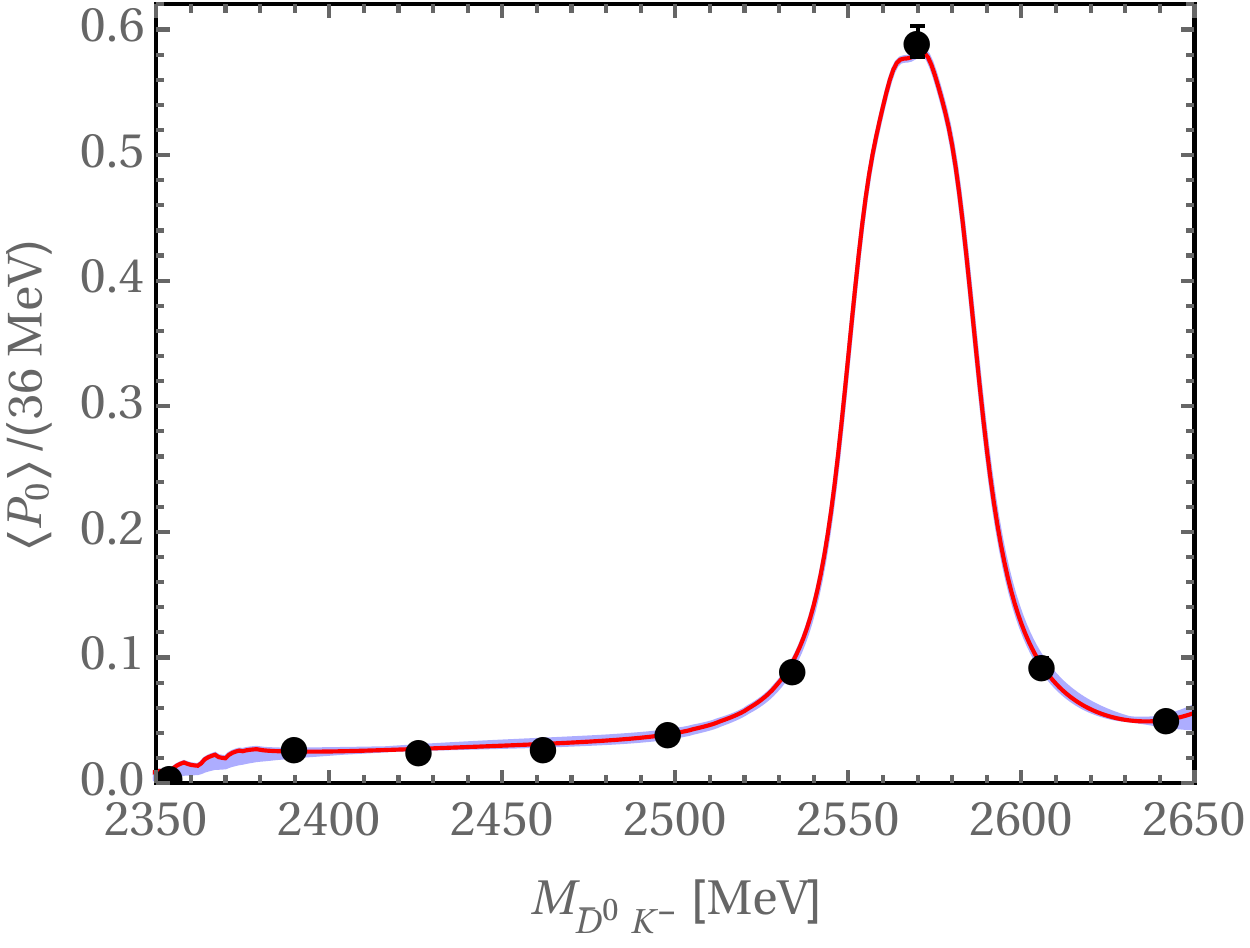} \hfill
  \includegraphics[width=0.32\linewidth]{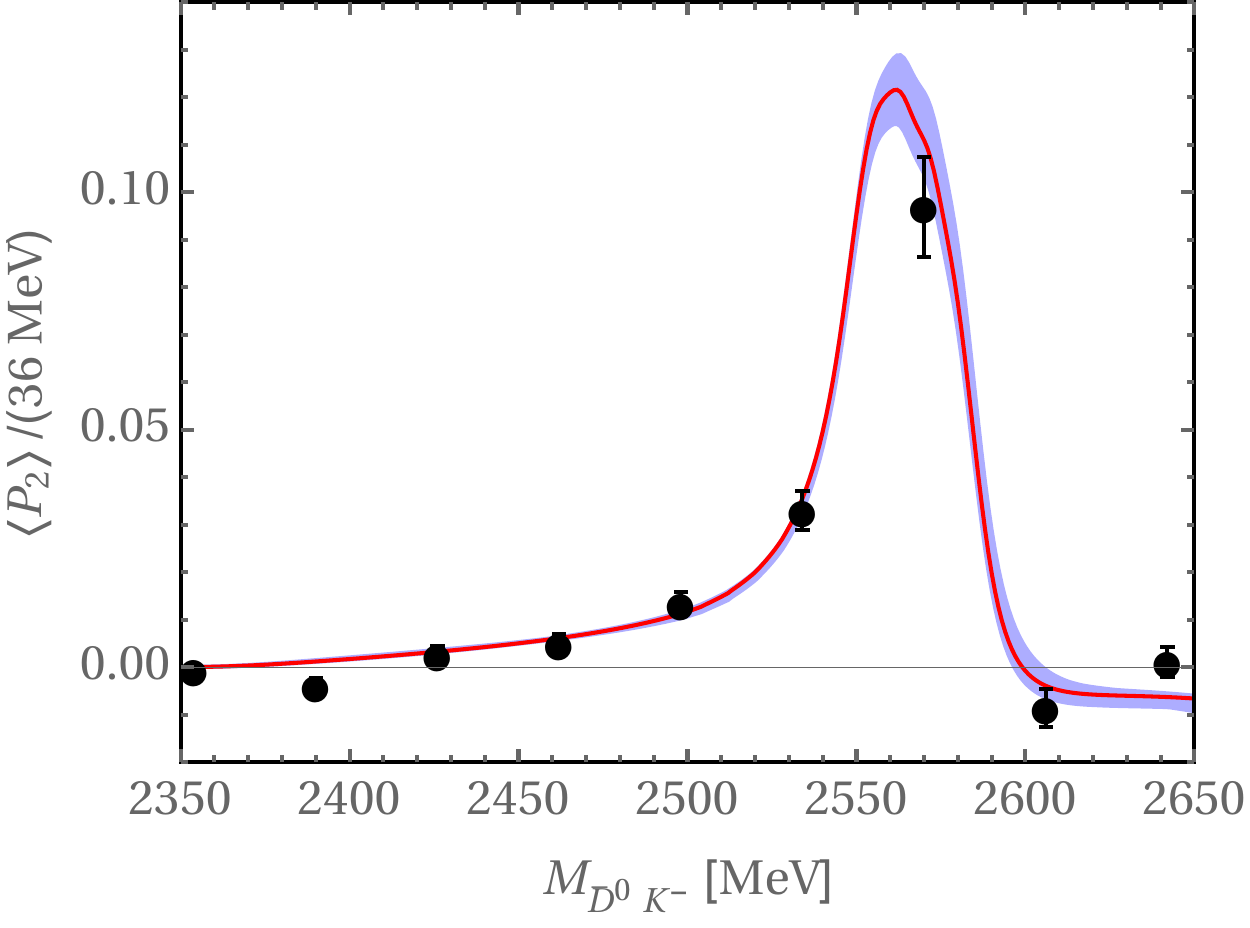} \hfill
  \includegraphics[width=0.31\linewidth]{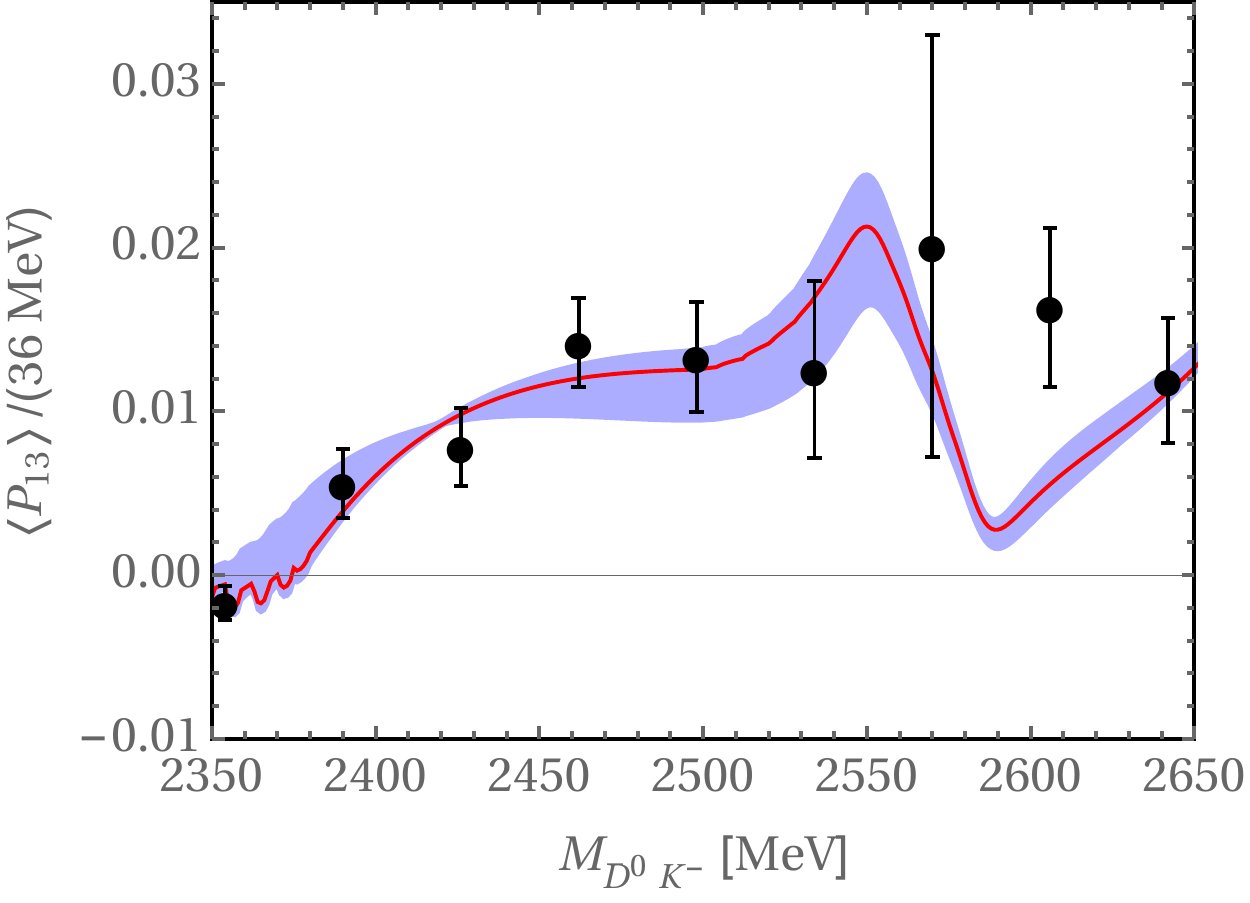}
 \end{center}
  \caption{
Fit to the LHCb data of the angular moments $\langle P_0\rangle$, $\langle P_2\rangle$ and $\langle P_{13}\rangle$
for the  $B_s^0\to \bar{D}^0K^-\pi^+$ reaction~\cite{Aaij:2014baa}. 
The largest error of $\langle P_1\rangle$ and $14\langle P_3\rangle/9$ in each bin is taken as
the error of $\langle P_{13}\rangle$. The error bands correspond to the one-sigma uncertainties propagated 
from the input scattering amplitudes.
\label{fig:fit2}}
\end{figure*}

\begin{figure*}[tb]
  \begin{center}
   \includegraphics[width=0.31\linewidth]{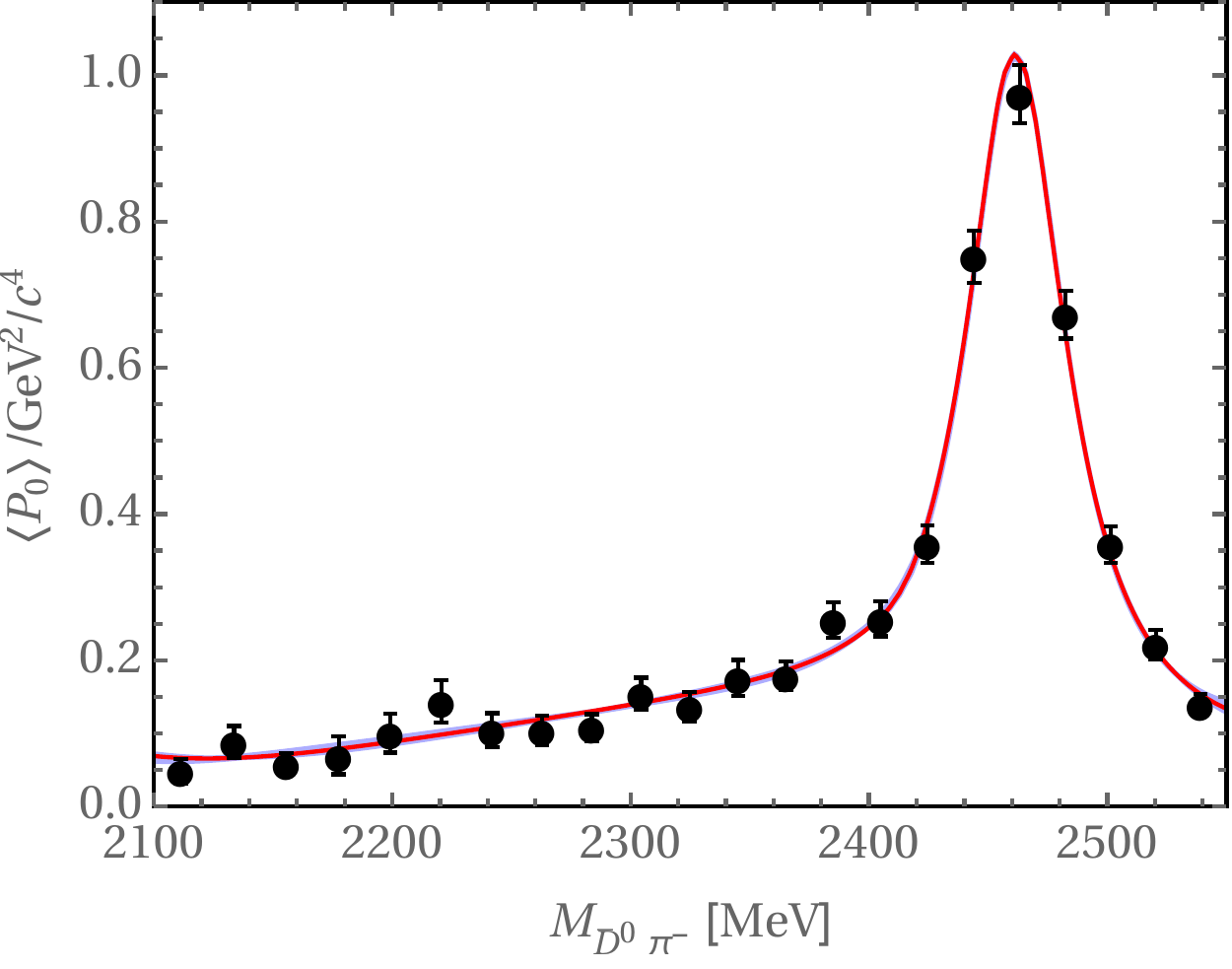} \hfill
  \includegraphics[width=0.32\linewidth]{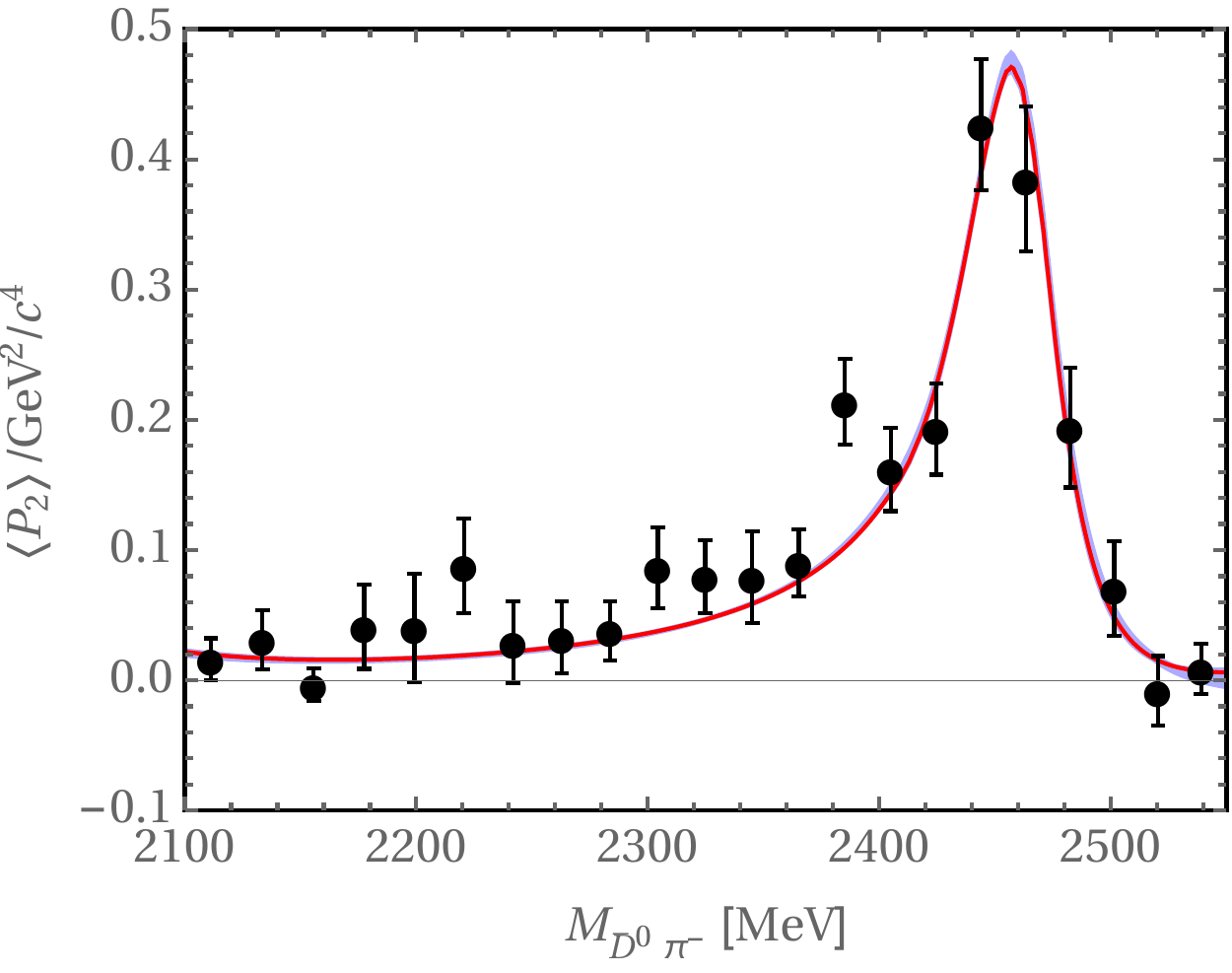} \hfill
  \includegraphics[width=0.31\linewidth]{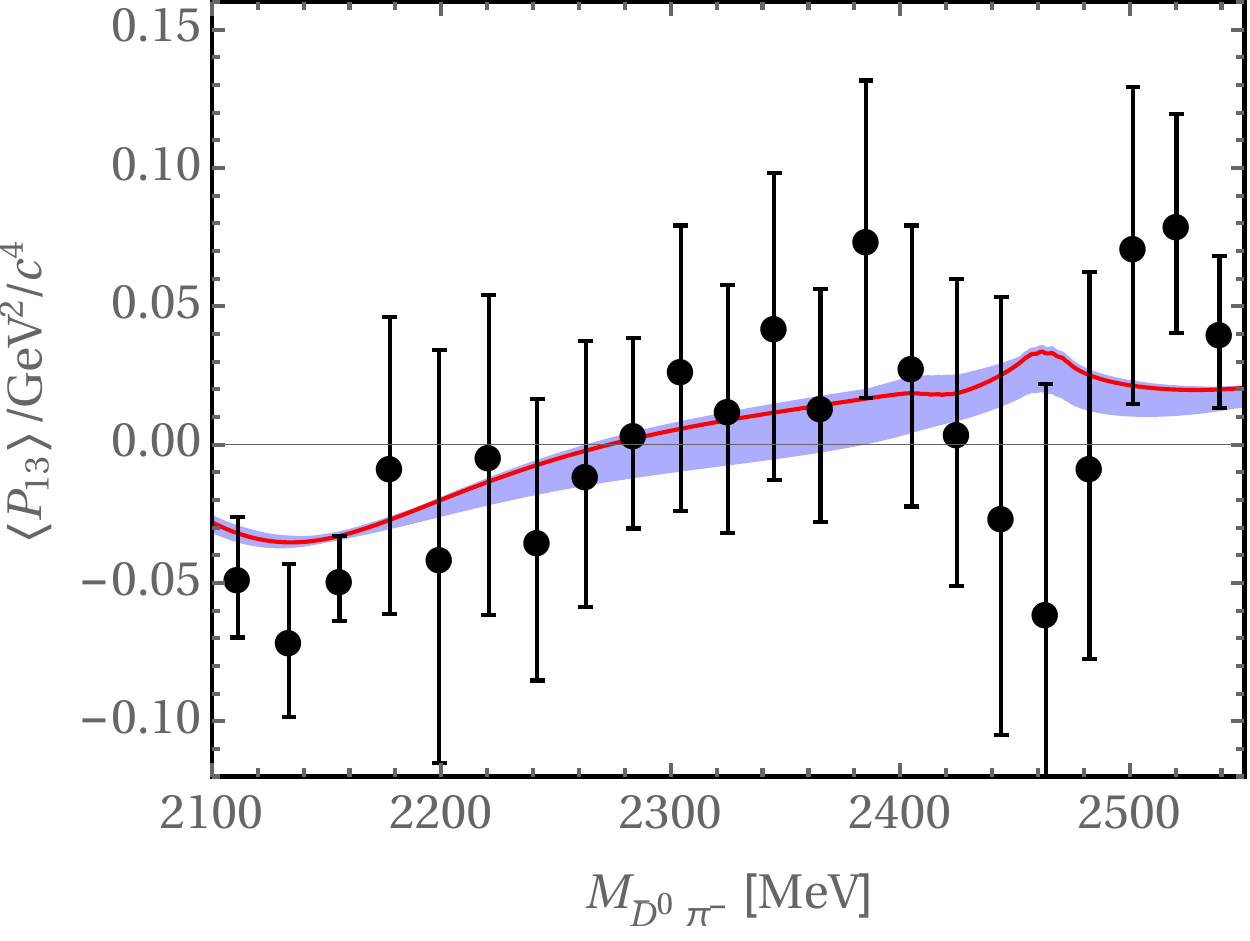}
 \end{center}
  \caption{
Fit to the LHCb data of the angular moments $\langle P_0\rangle$,
$\langle P_{13}\rangle$ and $\langle P_2\rangle$ for the $B^0\to \bar{D}^0\pi^-\pi^+$ reaction~\cite{Aaij:2015sqa}. 
The largest error of $\langle P_1\rangle$ and $14\langle P_3\rangle/9$ in each bin is taken as
the error of $\langle P_{13}\rangle$. The error bands correspond to the one-sigma uncertainties propagated 
from the input scattering amplitudes.
\label{fig:fit3}}
\end{figure*}

Moreover, we notice that only three sets of the weak production vertices in Table~\ref{tab:amps} are independent.
Thus instead of fitting the four decay amplitudes to the experimental angular moments simultaneously, 
we fix the LECs in the amplitudes in Eqs.~(\ref{eq:amp:1}-\ref{eq:amp:4}) by fitting to three of them, i.e.
$B^-\to D^+\pi^- K^-$, $B_s^0\to \bar{D}^0K^-\pi^+$ and $B^0\to \bar{D}^0\pi^-\pi^+$, and then describe the
angular moments for $B^0\to \bar{D}^0 \pi^- K^+$ with the determined LECs. We fit to the data of the moments
$\langle P_0\rangle$, $\langle P_{13}\rangle$, and $\langle P_2\rangle$ defined in Eqs.~\eqref{eq:angularm}
and \eqref{eq:angularm13} up to $M_{D\pi}=2.54$~GeV as in Ref.~\cite{Du:2017zvv} for the decays $B^-\to D^+\pi^- K^-$ and
$B^0\to \bar{D}^0\pi^-\pi^+$, and up to $M_{\bar D\bar K}=2.65$~GeV for $B_s^0\to \bar{D}^0K^-\pi^+$. The best fit results to the LHCb
data of the angular moments for the reactions $B^-\to D^+\pi^- K^-$, $B_s^0\to \bar{D}^0K^-\pi^+$ and
$B^0\to \bar{D}^0\pi^-\pi^+$ are shown in Figs.~\ref{fig:fit1}, \ref{fig:fit2} and \ref{fig:fit3}, respectively. The best fit
has a reasonable quality with $\chi^2/\text{d.o.f.}=1.2$ and the parameter values are listed in Tables~\ref{tab:fitLECs} and \ref{tab:fitsub}.
The bands in Figs.~\ref{fig:fit1}-\ref{fig:fit3} reflect the one-sigma errors of the parameters in the
scattering amplitudes determined in Ref.~\cite{Liu:2012zya}.

\begin{table}[tbh]
\caption{Values of the LECs from fitting to the LHCb data of the angular moments $\langle P_0\rangle$,
$\langle P_2\rangle$ and $\langle P_{13}\rangle$ for the reactions $B^-\to D^+\pi^- K^-$, $B_s^0\to \bar{D}^0K^-\pi^+$
and $B^0\to \bar{D}^0\pi^-\pi^+$ \cite{Aaij:2015vea,Aaij:2014baa,Aaij:2015sqa}. The two errors correspond to the
one-sigma uncertainties propagated from the input scattering amplitudes and from fitting to the experimental data, respectively.
The asterisk marks an input value from Ref.~\cite{Du:2017zvv}.}\label{tab:fitLECs}
\vspace{-0.5cm}
\bea
\begin{array}{c|c|c|c|c} 
\hline 
B/A^\ast & C/A & D/A & E/A & A^\prime/A \\
\hline
-3.6\pm0.4\pm0.1 & -2.3^{+1.3+0.5}_{-0.5-0.7} & 2.0^{+2.7+1.1}_{-1.0-1.1} & -1.6^{+1.0+1.2}_{-0.3-1.2} & 4.2^{+0.9+1.0}_{-2.2-0.8}\\
\hline\hline
B^\prime/A &C^\prime/A & D^\prime/A &E^\prime/A & \, \\
\hline
2.4^{+1.3+1.3}_{-6.5-1.3} & 1.4^{+7.0+1.7}_{-2.2-1.7} & -3.4^{+0.8+0.3}_{-0.5-0.3} & 2.4^{+3.9+1.1}_{-1.0-1.5} & \, \\
\hline 
\end{array}\nonumber
\eea
\end{table}

\begin{table}[tbh]
\caption{Values of the subtraction constant $a_A$ and the phase parameters, denoted by $\delta_{D^\ast}$,
$\delta_{D^\ast}^\prime$, and $\delta_{D_2}$. The two errors correspond to the uncertainties propagated from
the input scattering amplitudes and experimental data.}\label{tab:fitsub}
\vspace{-0.5cm}
\bea
\begin{array}{c|c|c|c|c} 
\hline 
\hline
 \, & a_A & \delta_{D^\ast} & \delta_{D^\ast}^\prime & \delta_{D_2} \\
 \hline
B^-\to D^+\pi^- K^-  & 0.4^{+0.1+0.2}_{-0.2-0.1} &  -1.2^{+0.1+0.2}_{-0.5-0.3} & -0.9^{+0.9+0.9}_{-0.2-0.7}  &-1.2^{+0.2+0.4}_{-0.9-0.4} \\
B_s^0\to \bar{D}^0K^-\pi^+ & -1.2^{+0.0+0.1}_{-0.2-0.0} & -2.9^{+0.4+0.1}_{-0.0-0.1} & - & -3.1^{+0.5+0.1}_{-0.1-0.1} \\
B^0\to \bar{D}^0\pi^-\pi^+ & -0.1^{+0.1+0.1}_{-0.4-0.2} & -0.3^{+0.3+0.3}_{-0.3-0.3} & - & -2.3^{+0.3+0.3}_{-0.2-0.3} \\
\hline
\end{array}\nonumber
\eea
\end{table}

\begin{figure*}[tbh]
  \begin{center}
   \includegraphics[width=0.33\linewidth]{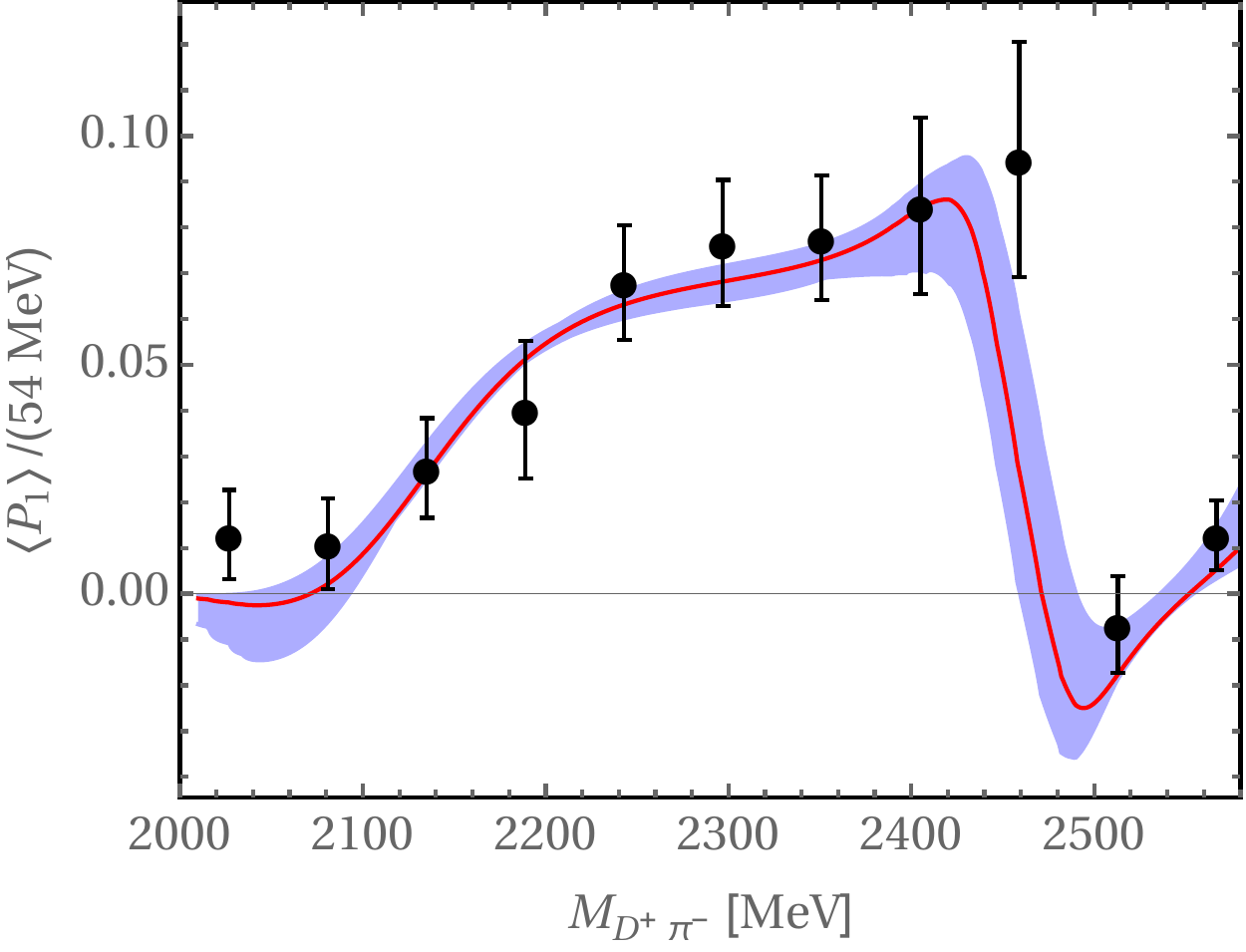} \qquad
  \includegraphics[width=0.33\linewidth]{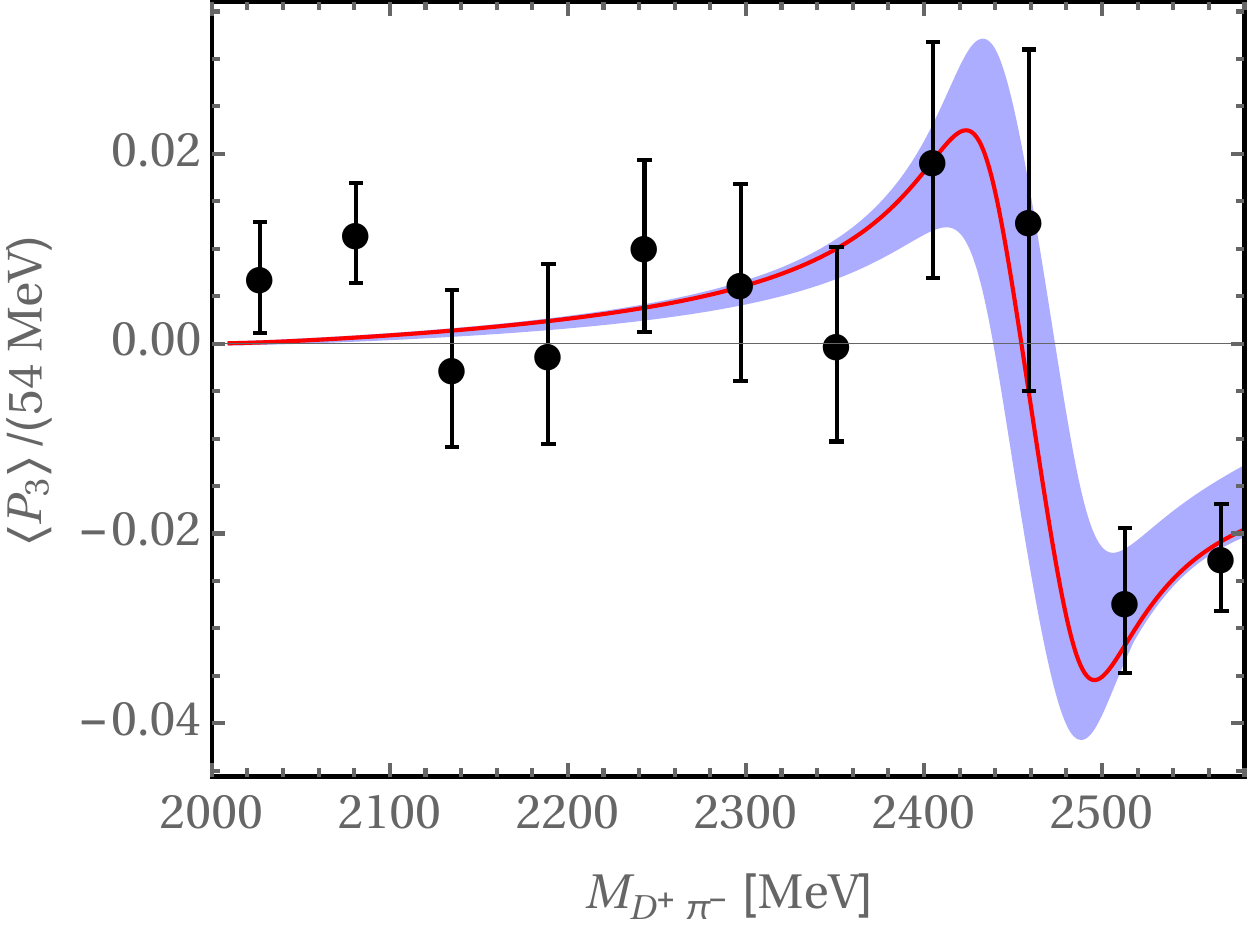}
 \end{center}
  \caption{Comparison of the predictions with the LHCb data of the angular moments
    $\langle P_1\rangle$ and $\langle P_3\rangle$ for the reaction $B^-\to D^+\pi^- K^-$~\cite{Aaij:2015vea}.
    The error bands correspond to the one-sigma uncertainties propagated from the input scattering amplitudes.
\label{fig:P13}}
\end{figure*}

\begin{figure*}[tbh]
  \begin{center}
   \includegraphics[width=0.33\linewidth]{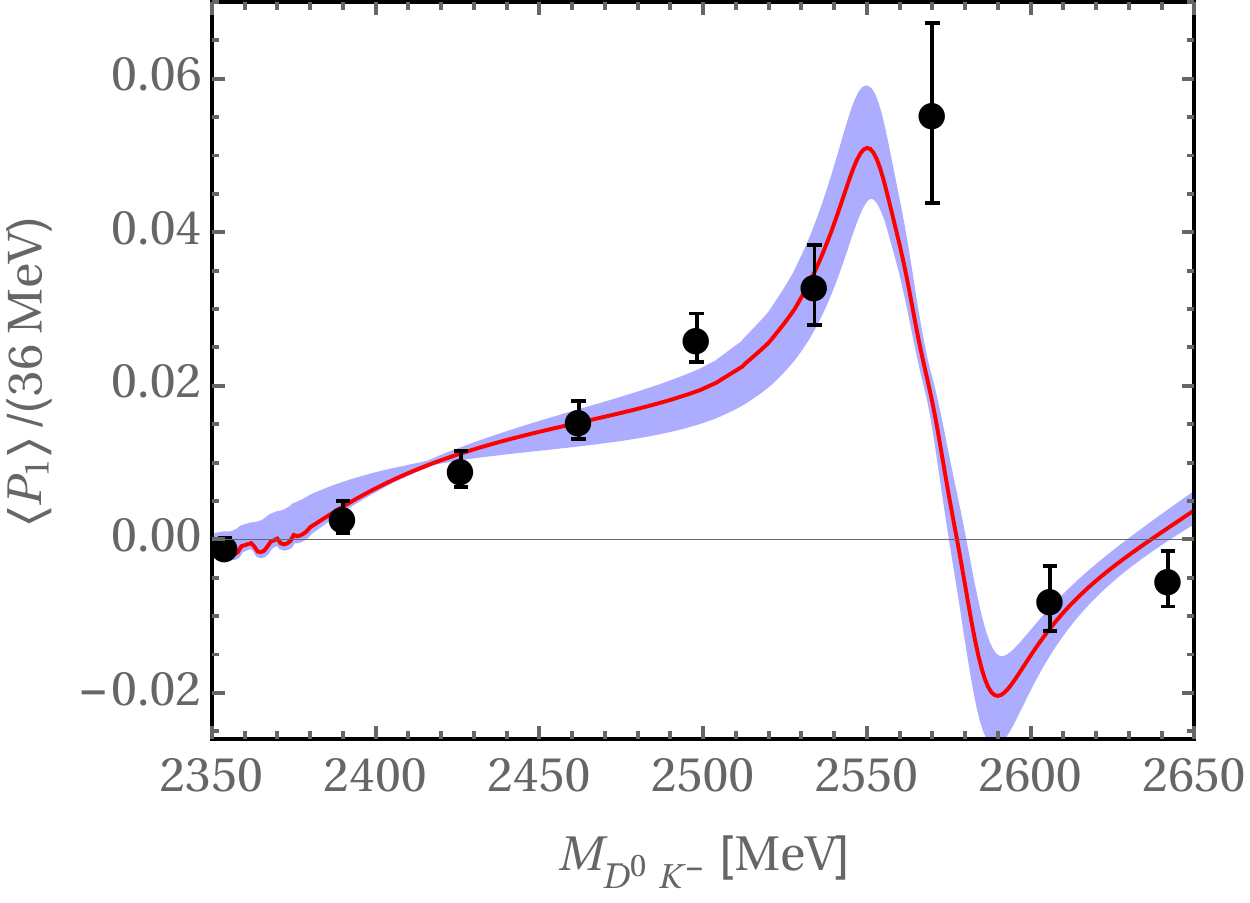} \qquad
  \includegraphics[width=0.33\linewidth]{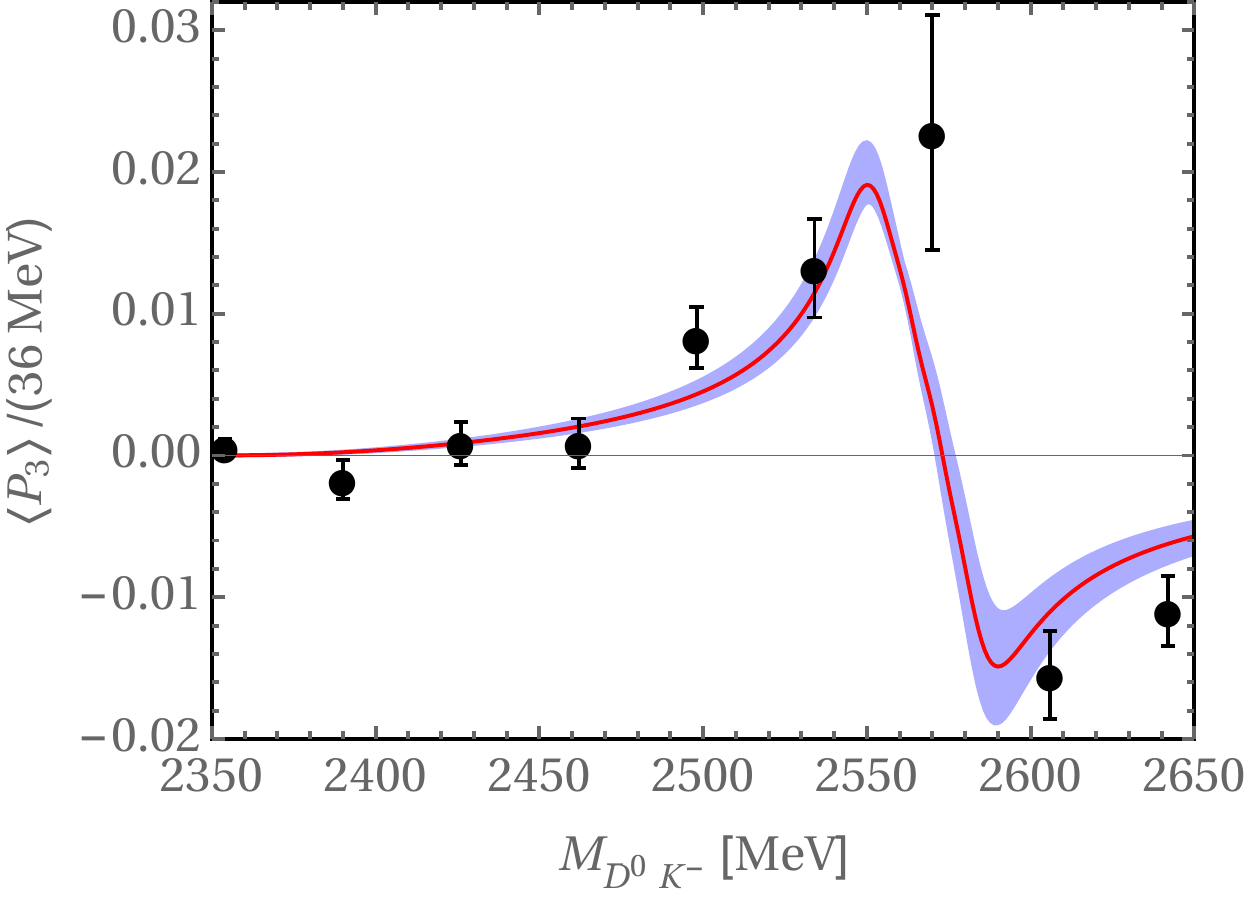}
 \end{center}
  \caption{Comparison of the predictions with the LHCb data of the angular moments
    $\langle P_1\rangle$ and $\langle P_3\rangle$ for the reaction $B_s^0\to \bar{D}^0K^-\pi^+$~\cite{Aaij:2014baa}.
    The error bands correspond to the one-sigma uncertainties propagated from the input scattering amplitudes.
\label{fig:P13bs0}}
\end{figure*}

\begin{figure*}[tbh]
  \begin{center}
   \includegraphics[width=0.33\linewidth]{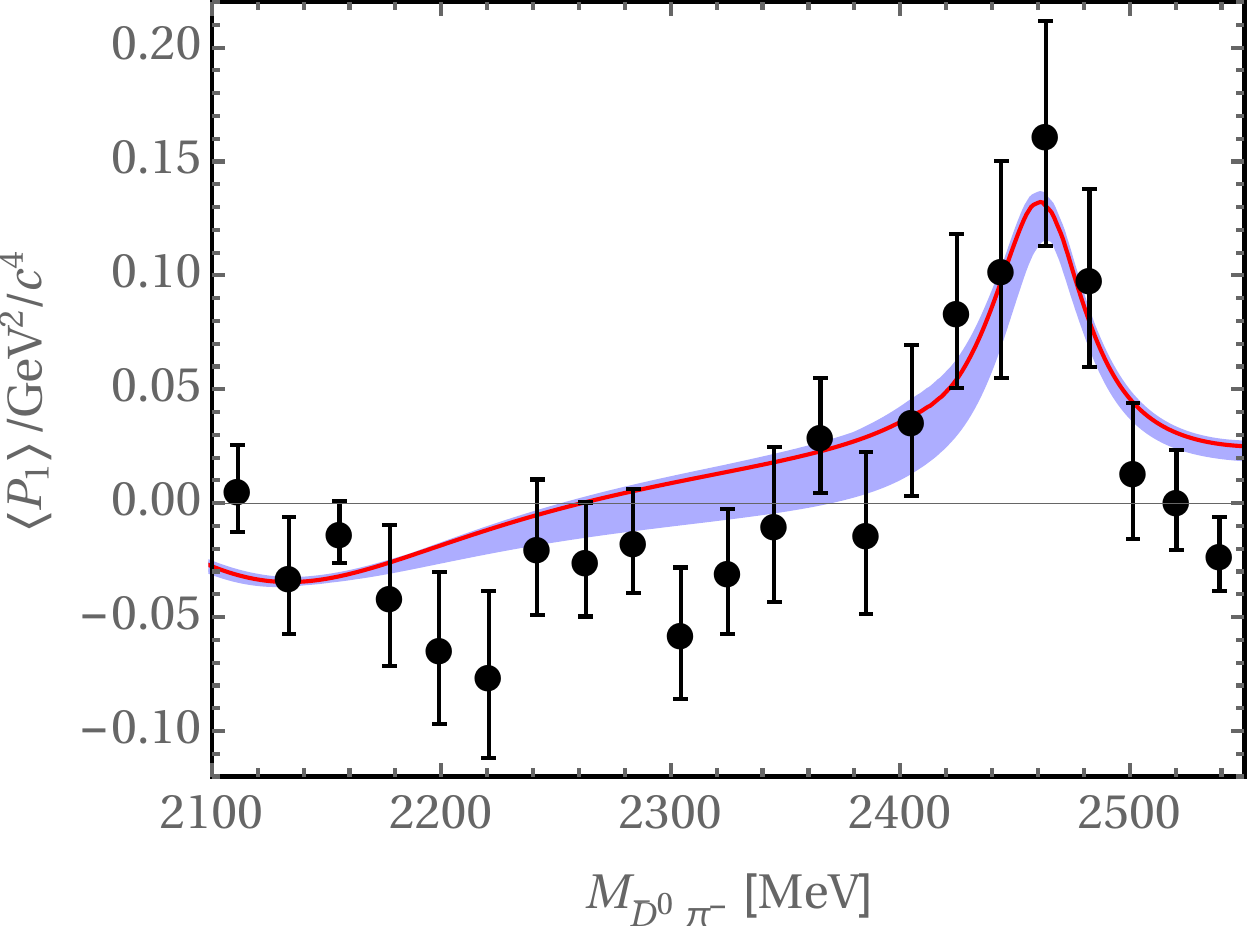} \qquad
  \includegraphics[width=0.33\linewidth]{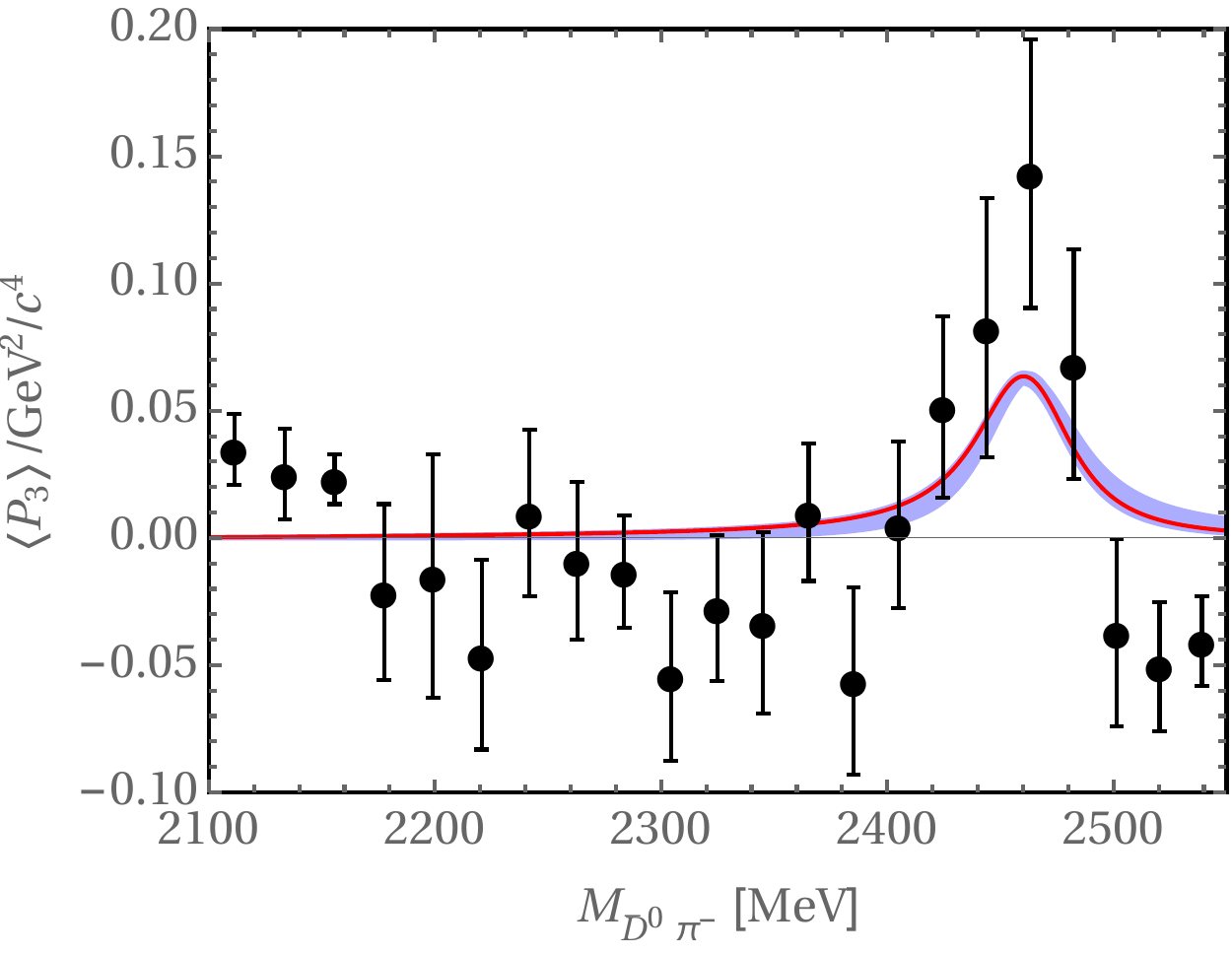}
 \end{center}
  \caption{Comparison of the predictions with the LHCb data for the angular moments
    $\langle P_1\rangle$ and $\langle P_3\rangle$ for the reaction $B^0\to \bar{D}^0\pi^-\pi^+$~\cite{Aaij:2015sqa}.
    The error bands correspond to the one-sigma uncertainties propagated from the input scattering amplitudes.
\label{fig:P13bb0}}
\end{figure*}

It is worth mentioning that in $\langle P_{13}\rangle$ for the reaction $B^-\to D^+\pi^- K^-$, where the
$D_2(2460)$ does not play any role, the data show significant variations between 2.4 and 2.5~GeV, as
that for $B^-\to D^+\pi^-\pi^-$ in Ref.~\cite{Du:2017zvv}. The data for the $B^0\to
\bar{D}^0\pi^-\pi^+$ and $B^0\to \bar{D}^0\pi^-K^+$ also have a similar behavior, but with lower statistics, see Fig.~\ref{fig:fit3} and Fig.~\ref{fig:fit4} below. 
These features are in line with the expectation due to the
opening of the $D\eta$ and $D_s\bar K$ thresholds, respectively, which leads to two cusps in the amplitude.
This also  implies the importance of the coupled-channel effects.
Moreover, using the parameters in Table~\ref{tab:fitLECs} and \ref{tab:fitsub}, we can predict the moments $\langle P_1\rangle$ and $\langle P_3\rangle$ separately, and the predictions are fully in line with the experimental
data~\cite{Aaij:2015vea,Aaij:2014baa,Aaij:2015sqa}, see Figs.~\ref{fig:P13}, \ref{fig:P13bs0} and \ref{fig:P13bb0}.

With the determined LECs in Table~\ref{tab:fitLECs}, we proceed to compare the angular moments for $B^0\to \bar{D}^0\pi^-K^+$  obtained from
the amplitudes in Eqs.~\eqref{eq:amp:4} and \eqref{eq:resonance2} with the LHCb data~\cite{Aaij:2015kqa}. Except for
the three normalization parameters (two for the resonances and one for the $S$-wave), there are only three
real free parameters: two phase factors for the resonances and one subtraction constant $a_A$ in the $G$ function
in the $S$-wave amplitude. We perform a fit in the energy region [2.1, 2.54]~GeV and the fit result is
shown in Fig.~\ref{fig:fit4}. The best fit has a $\chi^2/\text{d.o.f.}=1.8$ and the parameter values are
$a_A= 0.4^{+0.8}_{-0.5}$, $\delta_{D^\ast}=-2.5^{+0.2}_{-0.2}$, and $\delta_{D_2}=-5.9^{+0.2}_{-0.2}$, where only
the errors from fitting to the experimental data are given. From Figs.~\ref{fig:fit1}-\ref{fig:fit4}, we see that the
amplitudes in Eqs.~\eqref{eq:amp:1}-\eqref{eq:amp:4} indeed can describe the experimental data collected by the LHCb
Collaboration~\cite{Aaij:2014baa,Aaij:2015vea,Aaij:2015kqa,Aaij:2015sqa} quite well. Since the poles in those scattering amplitudes in the complex energy plane correspond to the scalar charmed meson resonances, one can conclude that the scalar charmed meson spectrum predicted in Refs.~\cite{Albaladejo:2016lbb,Du:2017zvv} is consistent with the LHCb data which are the best available data for the scalar charmed mesons. In particular, two poles, corresponding to two different resonances, exist in the $I=1/2$ nonstrange channel. The analyses in this work and in Ref.~\cite{Du:2017zvv} provide a strong support to such a spectrum, and thus the broad $D_0^\ast(2400)$ listed in RPP should be replaced by two states as advocated in Refs.~\cite{Albaladejo:2016lbb,Du:2017zvv}. The lower one has a mass of around 2.1~GeV and is much lighter than that extracted from experiments using a naive single-channel BW parameterization.

\begin{figure*}[tbh]
  \begin{center}
   \includegraphics[width=0.31\linewidth]{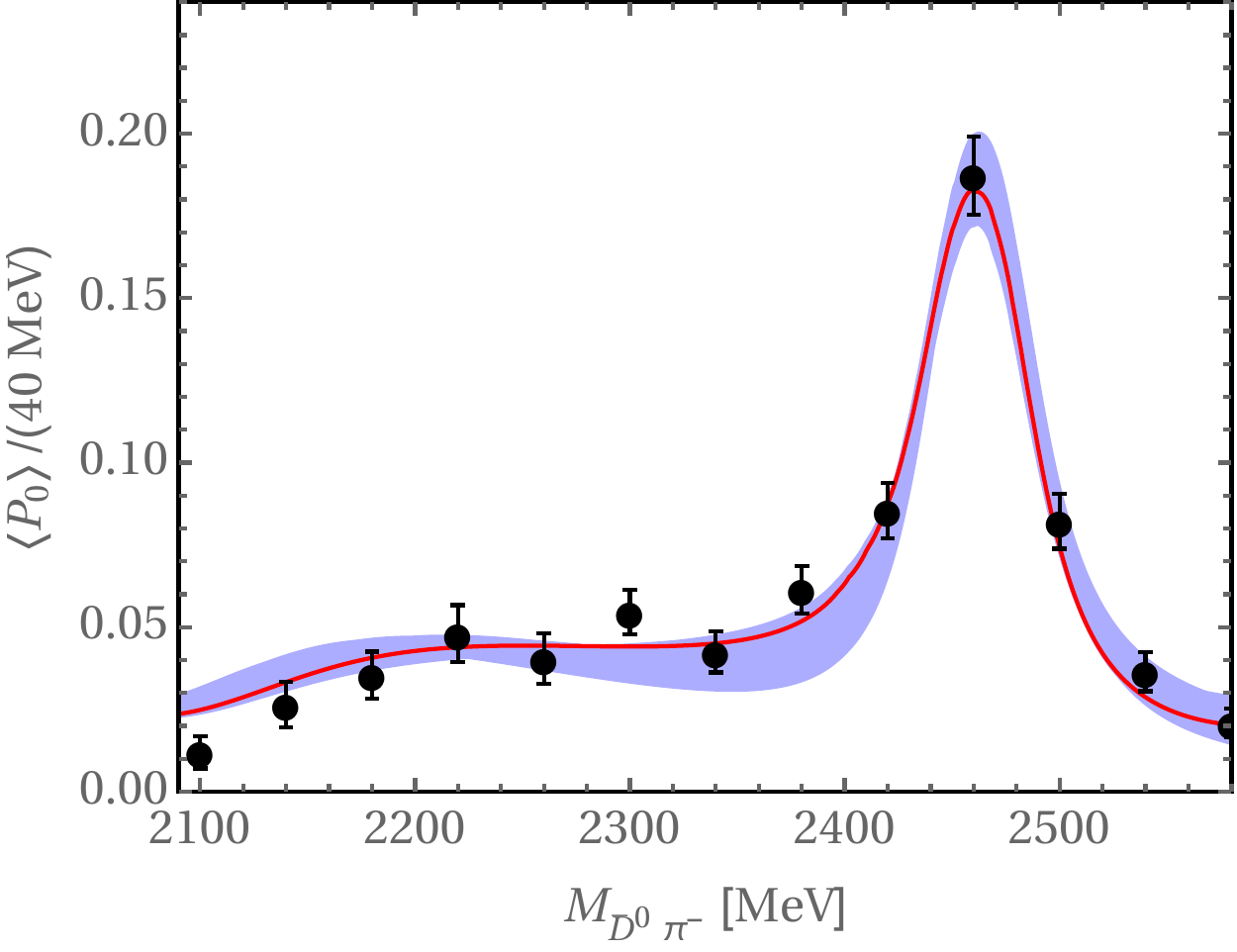} \hfill
  \includegraphics[width=0.32\linewidth]{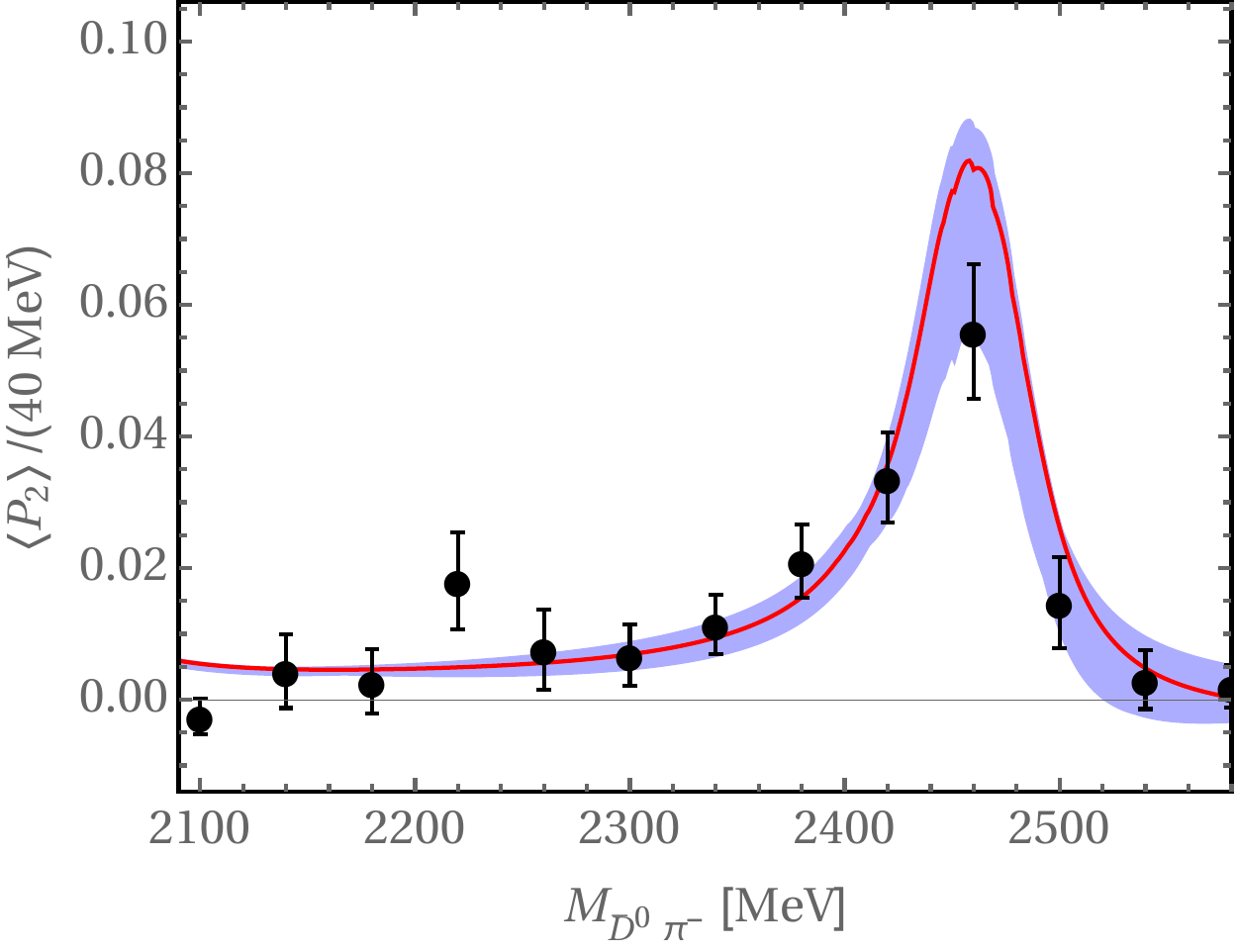} \hfill
  \includegraphics[width=0.31\linewidth]{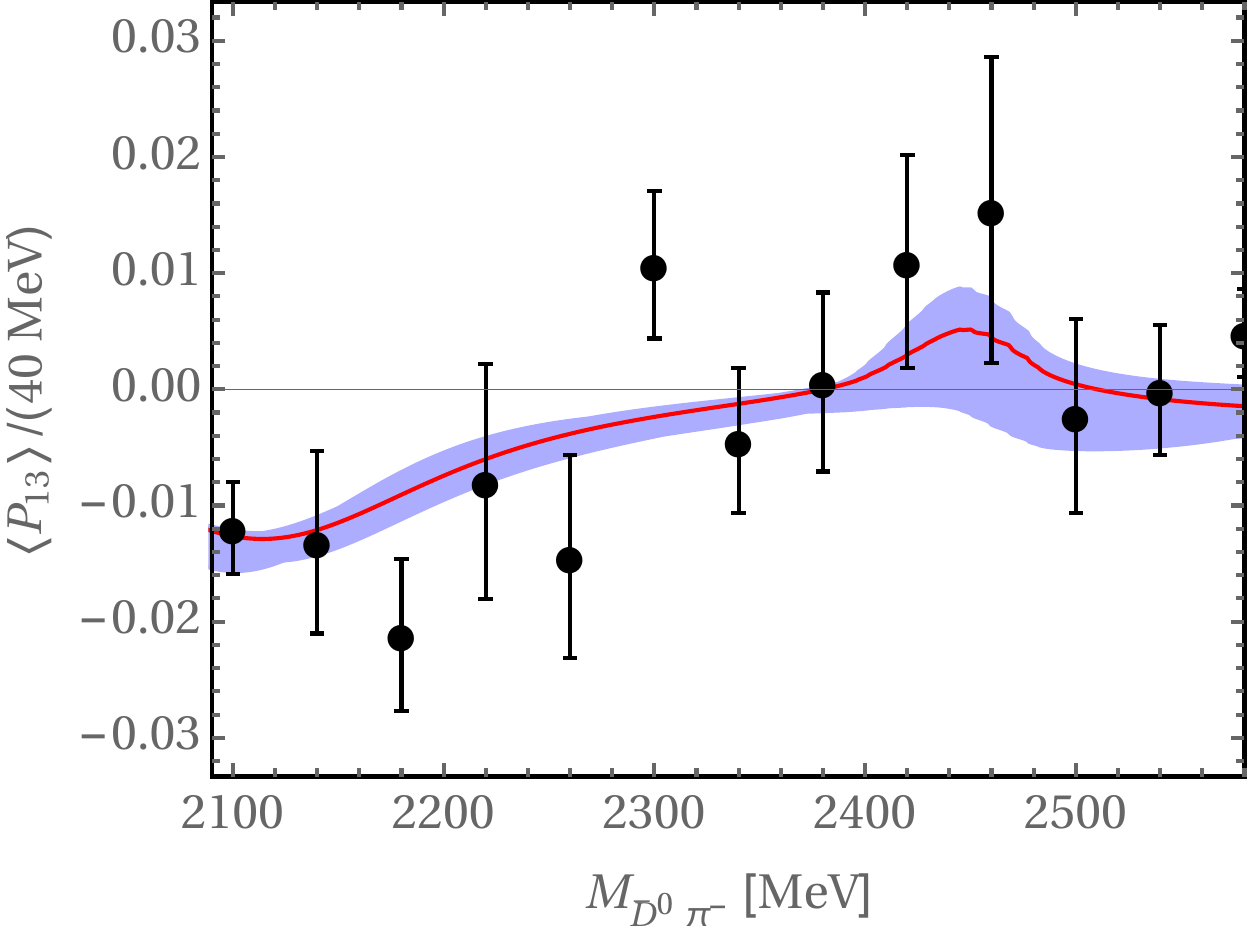}
 \end{center}
  \caption{
    Fit to the LHCb data of the angular moments for the  $B^0\to \bar{D}^0\pi^-K^+$ reaction~\cite{Aaij:2015kqa}.
    The shape of the $S$-wave is determined by only one real parameter $a_A$. The error bands correspond to the
    one-sigma uncertainties propagated from the input scattering amplitudes.
\label{fig:fit4}}
\end{figure*}

\section{Summary}\label{sec:sum}

We have shown that the constraints from  chiral symmetry are not fulfilled by the standard BW
parameterization for a pionic resonance. Using the BW parameterization to fit to the data can lead to a mass larger than its real value. In the case
that the resonance width $\Gamma$ is relatively narrow, one can show that the induced shift is proportional to $\Gamma^2$. Nevertheless, a naive modification of the BW parameterization
by introducing a pion energy factor is neither practical nor systematic since it can only be applied in a very limited energy region
before the coupled-channel effects become sizable. Fortunately, unitarized ChPT provides a framework
with the chiral symmetry constraint built in and coupled-channel effects taken into account. In the energy regime not far from the threshold of a charmed meson and a light pseudoscalar meson, it can thus be used to analyze the nonleptonic $B$ meson
three-body decays to charm mesons with the emission of two light pseudoscalar mesons which provide access to
the scalar charmed meson spectrum. 

In order to consider in detail the FSIs between the charm meson and the light
pseudoscalar meson for the reactions $B^-\to D^+\pi^-\pi^-$, $B_s^0\to \bar{D}^0K^-\pi^+$, $B^0\to\bar{D}^0\pi^-\pi^+$,
$B^-\to D^+\pi^-K^-$ and $B^0\to\bar{D}^0\pi^-K^+$, we have constructed a chiral effective Lagrangian for the decays of the
$B$- and $B_s$-mesons into a charmed meson associated with the emission of two light pseudoscalar mesons. Considering two-body
coupled-channel unitarity, we have constructed
the $D\phi$ $S$-wave amplitudes for these decays by making use of the effective weak Lagrangian. The involved LECs are thus free parameters in this framework. We have performed fits to the angular moments reported by the
LHCb Collaboration for a series of related decays $B_s^0\to \bar{D}^0K^-\pi^+$, $B^0\to\bar{D}^0\pi^-\pi^+$, $B^-\to D^+\pi^-K^-$, and  $B^0\to \bar{D}^0\pi^- K^+$~\cite{Aaij:2015vea,Aaij:2014baa,Aaij:2015sqa,Aaij:2015kqa}. In particular, the linear combination of two angular moments $\langle P_{13}\rangle = \langle P_1\rangle -{14}\langle P_3\rangle/9$ only depends on the $S$-$P$ interference when partial waves with $L\geq3$ are neglected. Thus, 
$\langle P_{13}\rangle$ is the quantity that one should focus on in order to learn about the scalar charmed mesons. We found that the LHCb angular moment data for all these decays can be well described. The predicted $\langle P_1\rangle$ and $\langle P_3\rangle$ also agree with the measurements.
Because the FSIs in these fits are taken from the unitarized ChPT amplitudes that were already fixed in Ref.~\cite{Liu:2012zya}, our analysis implies that the poles contained in these amplitudes can be regarded as the low-lying scalar charmed meson spectrum, and they such a spectrum is consistent with the best  data~\cite{Aaij:2014baa,Aaij:2015sqa,Aaij:2015vea,Aaij:2015kqa,Aaij:2016fma} in that sector. In particular, the poles of the scalar charm-nonstrange mesons, $\left(2105^{+6}_{-8}- i\, 102^{+10}_{-11}\right)$~MeV and $\left(2451^{+35}_{-26}-i\,134^{+7}_{-8}\right)$ MeV~\cite{Du:2017zvv}, are different from the resonance parameters of the $D_0^\ast(2400)$ listed in RPP~\cite{Tanabashi:2018oca}, which were extracted using a simple BW parameterization. The analysis in this work gives a further strong support to the two-$D_0^*$ scenario as advocated in Refs.~\cite{Albaladejo:2016lbb,Du:2017zvv}. Better knowledge of the scalar and axial-vector charmed mesons as well as the corresponding bottom counterparts are expected with the Belle-II~\cite{Kou:2018nap} and the upcoming high-luminosity and high-energy LHC~\cite{Cerri:2018ypt}.

\begin{acknowledgements}

This work is supported in part by the Deutsche Forschungsgemeinschaft (DFG)  and the National Natural
Science Foundation of China (NSFC) through the funds provided to the Sino-German Collaborative Research
Center ``Symmetries and the Emergence of Structure in QCD"  (NSFC Grant No. 11621131001, DFG Grant No. TRR110),
by the NSFC under Grant No. 11747601 and No. 11835015, by the Chinese Academy of Sciences (CAS) under Grant
No. QYZDB-SSW-SYS013 and No. XDPB09, by the CAS Center for Excellence in Particle Physics (CCEPP), by
the CAS President’s International Fellowship Initiative (PIFI) (Grant No. 2018DM0034), and by the VolkswagenStiftung (Grant No. 93562).

\end{acknowledgements}

\end{document}